\documentclass[twocolumn]{article} 

\usepackage[top=1.5cm, bottom=1.5cm, left=1.5cm, right=1.5cm,columnsep=0.5cm, includefoot]{geometry}

\usepackage{authblk}
\usepackage{graphicx}
\usepackage{amsmath}
\usepackage{amsfonts}
\usepackage{cite}
\usepackage{url}

\graphicspath{{./figs/}}
\def\vec#1{\mathbf{#1}}
\def\bsym#1{\boldsymbol{#1}}
\def\i.e.{\textit{i.e.}}
\def\e.g.{\textit{e.g.}}

\title{Guided Facial Skin Color Correction}

\author[1]{Keiichiro Shirai}
\author[2]{Tatsuya Baba}
\author[3]{Shunsuke Ono}
\author[4]{Masahiro Okuda}
\author[1]{Yusuke Tatesumi}
\author[5]{Paul Perrotin}

\affil[1]{Shinshu University, Japan}
\affil[2]{The University of Kitakyushu}
\affil[3]{Tokyo Institute of Technology, Japan}
\affil[4]{Doshisha University, Japan}
\affil[5]{The University Institute of Technology La Rochelle, France}

\date{}

\begin{document}
\maketitle

\begin{abstract}
This paper proposes an automatic image correction method for portrait photographs, which promotes consistency of facial skin color by suppressing skin color changes due to background colors. In portrait photographs, skin color is often distorted due to the lighting environment (\e.g., light reflected from a colored background wall and over-exposure by a camera strobe), and if the photo is artificially combined with another background color, this color change is emphasized, resulting in an unnatural synthesized result. 
In our framework, after roughly extracting the face region and rectifying the skin color distribution in a color space, we perform color and brightness correction around the face in the original image to achieve a proper color balance of the facial image, which is not affected by luminance and background colors. Unlike conventional algorithms for color correction, our final result is attained by a color correction process with a guide image. 
In particular, our guided image filtering for the color correction does not require a perfectly-aligned guide image required in the original guide image filtering method proposed by He \textit{et al}. 
Experimental results show that our method generates more natural results than conventional methods on not only headshot photographs but also natural scene photographs. We also show automatic yearbook style photo generation as an another application. 
\end{abstract}


\section{Introduction}

Portrait photographs sometimes get undesirable color due to a background color reflection, so that we need to make the skin color and brightness uniform for each photograph. In such a situation, a professional cameraman usually arranges lighting conditions with special equipment.
Meanwhile, for an amateur, more effort is required to achieve the same quality.
This leads to a demand for an automatic and simple method that does not require any special equipment for unifying the color and brightness between multiple images.
The undesirable color on portrait photographs is caused by the following. 
\begin{itemize}
\setlength{\parskip}{1mm}
\setlength{\itemsep}{0mm}
\item Each camera has its own camera response sensitivity. Thus, the distribution of skin color in a color space also depends on the camera used.
\item Background color is usually reflected to faces resulting in distorted colors.
\item If each subject wears clothes with more than one color, color correction for the whole image region distorts the skin color, and skin color correction will discolor the clothes. 
\end{itemize}
Therefore, a color correction scheme that affects only the face region is required.

In this paper, we attempt to correct an image with various color and brightness conditions with a guide image. 
Our method has various applications, \textit{e.g.}, the creation of a photo name list such as a yearbook with high quality facial images without much effort or cost. If one tries to gather photographs taken in different lighting conditions with different cameras, correcting the images becomes a very difficult task even using off-the-shelf image processing software.
In the situation considered, we need to unify several attributes of all images: the size of cropped facial image, facial skin color, brightness, and background color.
Automatic correction of the attributes requires
\begin{enumerate}
\renewcommand{\labelenumi}{(\arabic{enumi}).}
\setlength{\parskip}{1mm}
\setlength{\itemsep}{0mm}
\item Face detection and Facial skin color extraction.
\item Facial skin color correction.
\end{enumerate}
A simple combination of these techniques gives boundary artifacts between the corrected region and the uncorrected region because the color correction only corrects the extracted region. This is described in Sec.~\ref{sec:relatedwork}.
%
%

With the algorithm described in this paper, correction of facial skin color and brightness is performed using a hybrid \textit{guided image filtering} (GIF) method.
Because region extraction and color transfer are individual procedures, color transfer on part of an image generates a color gap between the original color part and the color transformed part.
For this problem, we propose the hybrid GIF (the red box in Fig.~\ref{fig:algorithm}), which performs color transfer and segmentation.
Only the face region is extracted (Fig.~\ref{fig:algorithm}(1)), followed by adjustment with grading, and then the hybrid GIF corrects the face region of the input image by using the corrected image as a guide image while keeping colors in other regions (Fig.~\ref{fig:algorithm}(2)). In other words, the hybrid GIF transforms a part of the image as an object based color transfer.
In contrast to the fact that the original GIF \cite{He2013} needs a perfectly-aligned guide image, our GIF only needs roughly aligned pixels. 
As for non-aligned regions, the filtering is achieved by aligned region propagation such as colorization \cite{Levin2004} and matting \cite{Levin2008}.
Our method carries out nonlinear correction only on the face region and achieves better results than conventional methods.
A preliminary version of this study without several improvements or new applications appeared in conference proceedings \cite{BB2015ACPR}.

The rest of this paper is organized as follows. Sec.~\ref{sec:relatedwork} introduces related work on color correction and facial image correction. Sec.~\ref{sec:proposed} describes the proposed guide facial skin color correction method. The proposed algorithm is evaluated in experiments by comparison with various conventional methods in Sec.~\ref{sec:results}.
Finally, this paper is concluded in Sec.~\ref{sec:conclusion}.

%
%
%

\section{Related work}\label{sec:relatedwork}

Reference based color correction is one of the essential research issues for image editting. \textit{Color grading} methods \cite{Pitie2007,HaCohen2011,HaCohen2013}, which match the color attributes (tone curve and color distribution for each color) of a target image with those of another target image, are effective for color correction among multiple images. However, these existing techniques are global image correction operators and assume that subjects wear the same kinds of clothes. 
Other related techniques are image \textit{matting} \cite{Levin2008,He2010}. By using these techniques, one can change colors by specifying desired colors to represent pixels. However, to obtain natural colorization, much coloring information is needed. 
Another possible approach is to apply color transfer based on GIF with guide images \cite{Petschnigg2004,Eisemann2004,Rabin2010,He2013}, where coloring information is given in a block as a guide image. However, the guide image used in the methods is limited to perfectly aligned images without any position gap, which is often not the case for the situation considered.

As for facial image correction methods, the authors of \cite{Batool2014} propose a wrinkle removal method but the purpose is different from color grading of facial skin color.
The method proposed in \cite{Shin2014} can also transfer the style of the original image to the target one. However, it also edits clothes and performs well only for headshot photos.

\section{Proposed method}\label{sec:proposed}

\begin{figure}[t]
\centering
\includegraphics[width=\linewidth]{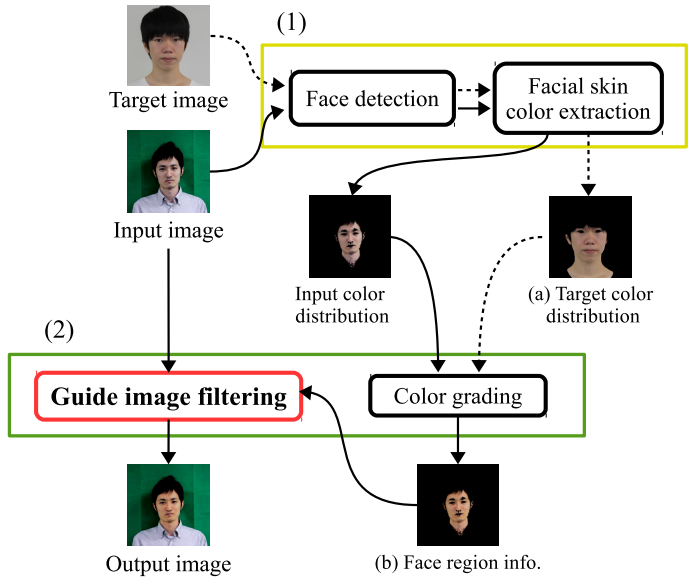} 
\caption{Flow chart of our method.}
\label{fig:algorithm}
\end{figure}

Our algorithm transforms colors of a part of an input image using a target image color unlike general color transfer methods and corrects surrounding colors of the transformed part.
Fig.~\ref{fig:algorithm} illustrates the flow chart of our method. 
Our method mainly consists of two parts.
\begin{enumerate}
\renewcommand{\labelenumi}{(\arabic{enumi}).}
\setlength{\parskip}{1mm}
\setlength{\itemsep}{0mm}
\item \textbf{Face detection and Facial skin color extraction} (the yellow box in Fig.~\ref{fig:algorithm}) : It detects the face part and extracts its facial skin color.
\item \textbf{Skin color correction} (green box) : After extracting the skin region\footnote{We distinguish \textit{region} and \textit{area}. \textit{Area} indicates the detected face area in Sec.~\ref{sec:face_detect}, while \textit{region} indicates the extracted facial skin color region in Sec.~\ref{sec:skin_extract}.} in Step (1), the distribution of the facial skin color is rectified using the image (a). The color of the face region is modified by using the image (b) as the guide image.
\end{enumerate}
Note that this paper uses the two phrases, \textit{skin color} and \textit{facial skin color}. \textit{Skin color} refers to the skin color of the whole body, while \textit{facial skin color} is the color of the skin in the face region only.

The novelty of the proposed method lies in (2), where the \textbf{color correction with grading affects only a part of the image}.
We use conventional methods for the other steps such as the face detection with some modifications. Each procedure is described in more detail hereafter.

\begin{figure}[t]
\centering
\setlength{\tabcolsep}{1mm}
\begin{tabular}{ccc}
\includegraphics[height=28mm]{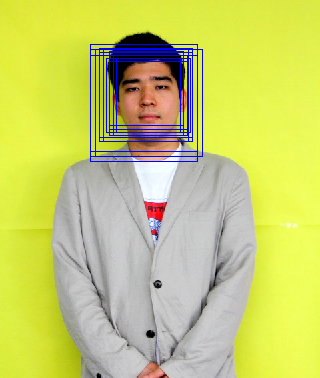} &
\includegraphics[height=28mm]{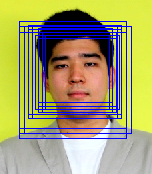} & 
\includegraphics[height=28mm]{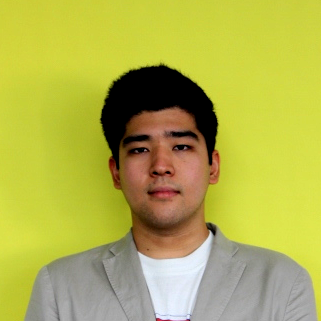}
\end{tabular}
\caption{Face detection. (left) Original image, (center) Detected face area (candidates), and (right) Face area in the original image. Blue boxes indicate candidates for the face.}
\label{fig:face_detect}
\end{figure}
%

%
%

\subsection{Face detection} \label{sec:face_detect}

In the first step, we detect an area from head to shoulder (Fig.~\ref{fig:face_detect}), where we use Haar-cascade detection in OpenCV, also known as the Viola-Jones algorithm \cite{Viola2001}, for the face detection.
The face area is described by $n$ rectangle windows, where $n$ is the number of candidates, with a set of barycentric coordinates $(x,y)$ and the sizes of the detected candidate rectangles $(w,h)$ Fig.~\ref{fig:face_detect} (center). The blue boxes in Fig.~\ref{fig:face_detect} indicate the detected candidate rectangles. We find the median of $x,y,w,h$ as
\begin{equation}
\arraycolsep=1mm
\begin{array}{rlrl}
\widehat{x} &:= \text{median}( \{x_1,\ldots,x_n\} ), \ & \widehat{w} &:= \text{median}( \{w_1,\ldots,w_n\} ), \\
\widehat{y} &:= \text{median}( \{y_1,\ldots,y_n\} ), \ & \widehat{h} &:= \text{median}( \{h_1,\ldots,h_n\} ),
\end{array}
\end{equation}
where the subscript is the index of the candidate.
The rectangle with the median values of the barycentric coordinate $(\widehat{x},\widehat{y})$  is adopted as the final face area. Since the size of the face area $(\widehat{w},\widehat{h})$ depends on the image, the size of the face area is adjusted by: 
\begin{equation}
\left(2l\,\widehat{w}+1\right)\times \left(2l\,\widehat{w}+1\right),
\label{eq:eq2}
\end{equation}
where $l$ is the prescribed scale factor.

%
%

\subsection{Skin color extraction} \label{sec:skin_extract}

Rough extraction of the skin color region near the face is performed by a common approach using morphological labeling in the image space and clustering in the color space:
\begin{enumerate}
\renewcommand{\labelenumi}{(\roman{enumi})}
\setlength{\parskip}{1mm}
\setlength{\itemsep}{0mm}
\item We classify the color of each pixel by clustering the color distribution of the entire image in the HSV color space, and for each pixel we allocate the label of the cluster that each pixel belongs to.
\item Some regions are generated by concatenating neighboring pixels with the same labels in the image space, and then we extract regions mainly in the detected face area $\Omega_\text{rect}$ (Eq.~\eqref{eq:eq2}) and assign them to the facial skin region.
\end{enumerate}
Note that, since the skin color in the detected face area (Sec.~\ref{sec:face_detect}) is known, we may obtain satisfactory results by using simpler approaches such as the
\textit{k-means} method \cite{Lloyd1982}
with stable seed cluster selection \cite{Arthur2007} (\textit{i.e.}, \textit{k-means}++) for clustering color distributions.

In this paper, for simplicity, we apply \textit{k-means} clustering \cite{Arthur2007} to the 1-D distribution of the \textit{hue} values of the entire image, and we perform the above-mentioned procedures (i) and (ii) to detect the face region $\Omega^\textit{hue}_{S}$. 
Then we perform thresholding on the region $\Omega^\textit{hue}_{S}$ for the \textit{saturation} and \textit{value} components of every pixel. 
We define the skin color region $\Omega_{S}$ and the condition for the skin color at pixel $p$ experimentally as follows:
%
%
%
\begin{equation}
\Omega_{S} := \left\{ p \mid p \!\in\! \Omega_{S}^{hue}, \ s_p \!\in\! [\widehat{s}\!-\!0.2, \widehat{s}\!+\!0.2],\ v_p \!\in\! [0.15, 0.95] \right\}
\end{equation}
where $\widehat{s}$ is the median of the \textit{saturation} in the face area. 
The \textit{saturation} and the \textit{value} are normalized to $[0,1]$ and the threshold values are set to excellently extract the facial skin color region on our dataset, which is available on the project page.
The number of clusters depends on the photographic environment. We set $k=4$ in this paper.

\begin{figure}[t]
\centering
\setlength{\tabcolsep}{.5mm}
\begin{tabular}{cccc}
\includegraphics[trim= 10 0 30 0,height=2.5cm,clip]{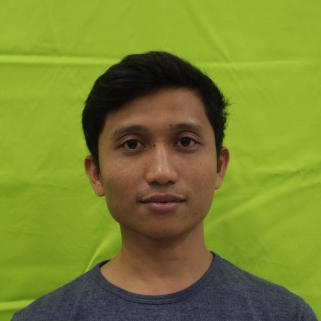}&
\includegraphics[trim= 10 0 30 0,height=2.5cm,clip]{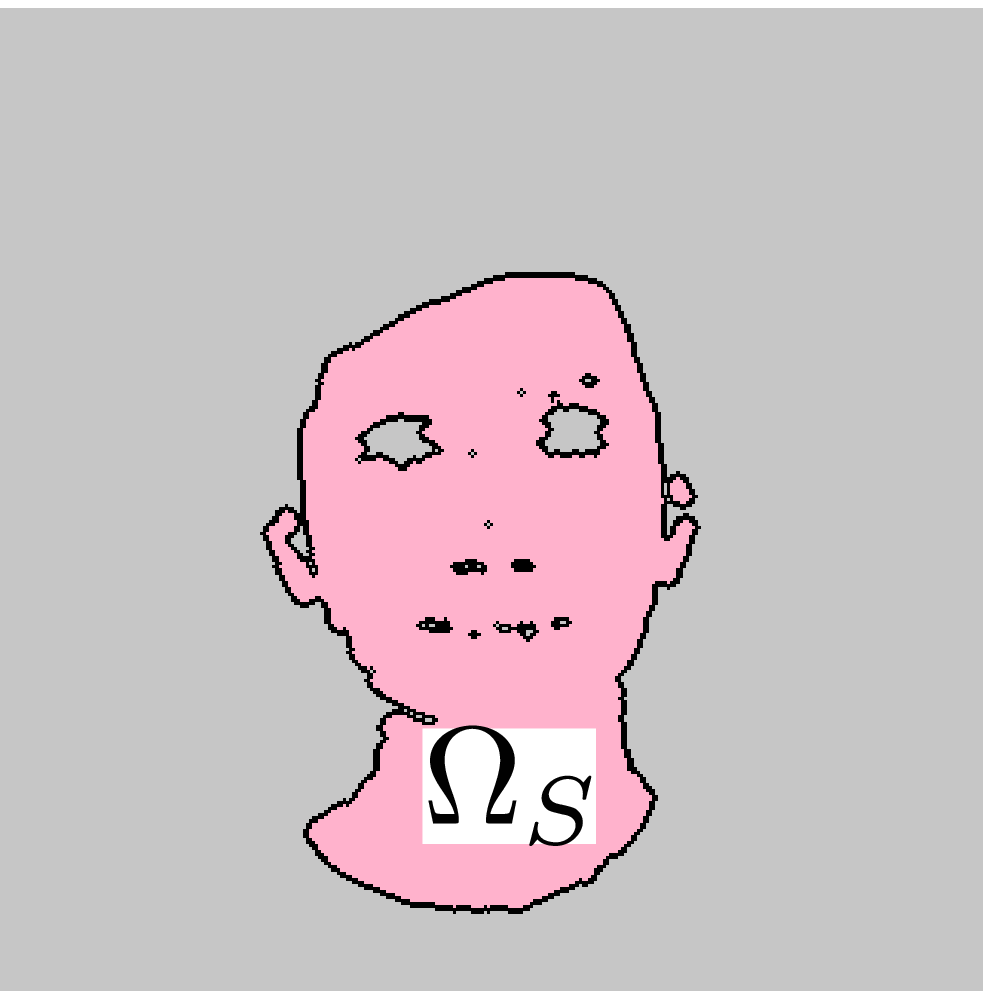}&
\includegraphics[trim= 10 0 30 0,height=2.5cm,clip]{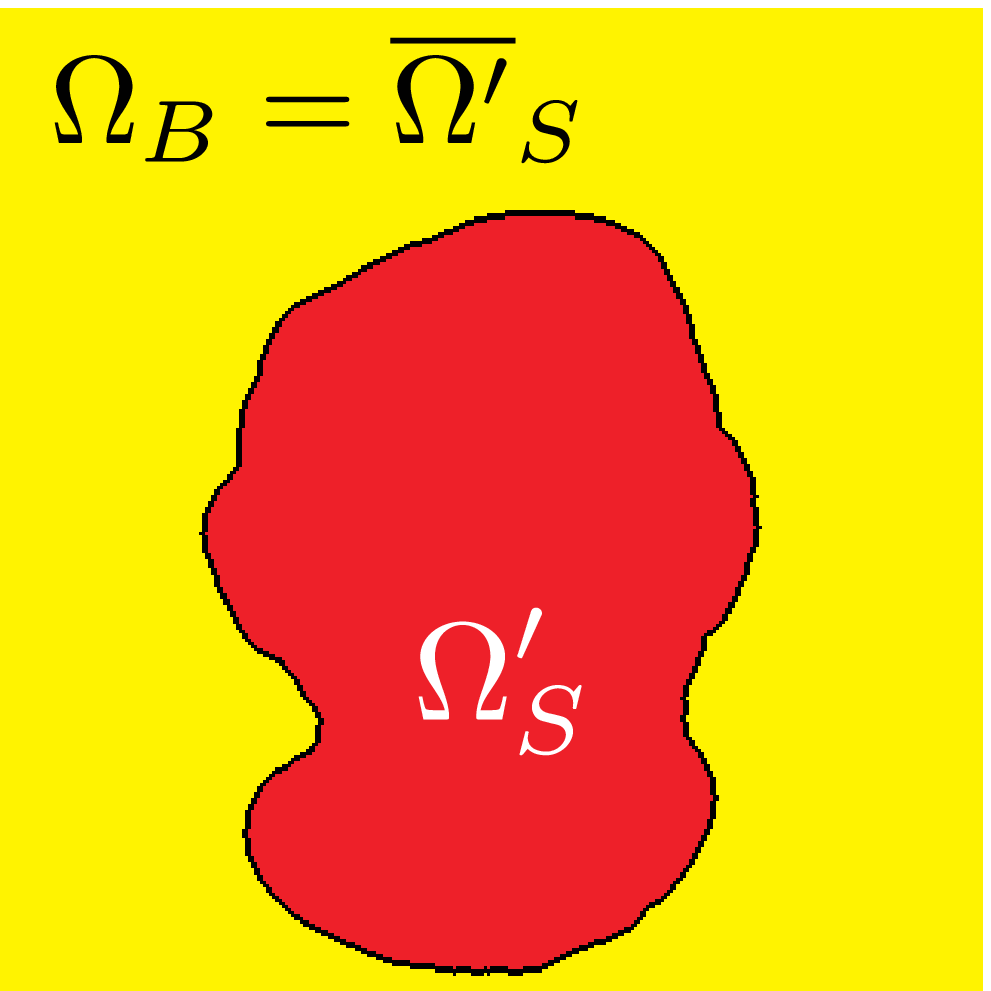}&
\includegraphics[trim= 10 0 30 0,height=2.5cm,clip]{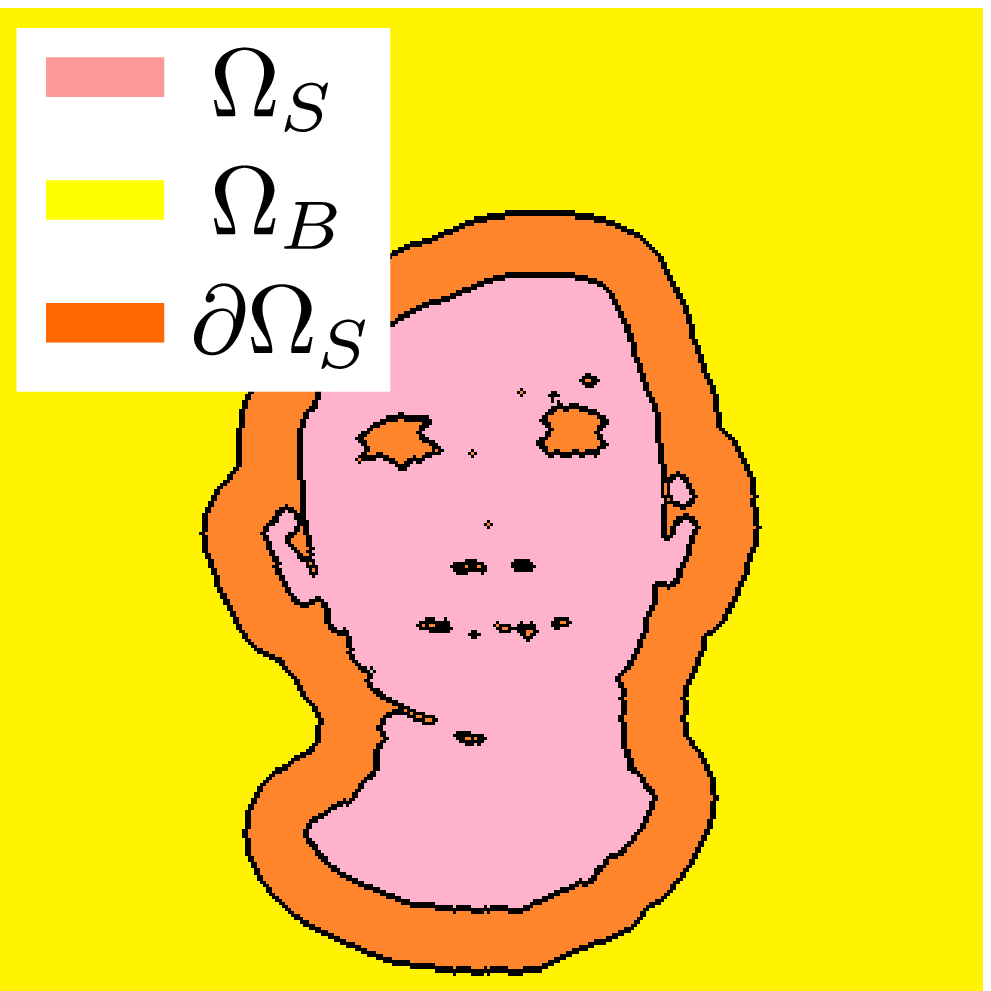}
\\
(a) & (b) & (c) & (d)
\end{tabular}
\caption{Region example. (a) Input image, (b) Skin color region $\Omega_{S}$, (c) Dilated skin color region $\Omega_{S}'$, and (d) Each region in the whole image. Pink, red, yellow, and orange indicate each region $\Omega_{S}$, $\Omega_{S}'$, $\Omega_{B}$, and $\Omega_{\partial S}$, respectively.}
\label{fig:eachregi}
\end{figure}

For our guide image filtering, we define $\Omega_{B}$ and $\Omega_{\partial S}$ using the dilation function, which is a type of morphological operation.
They are given as
\begin{equation}
\begin{aligned}
\Omega_{S}' &:= \text{dilation}(\Omega_{S}),\\
\Omega_{B} &:=  \overline{\Omega'}_{S},\\
\Omega_{\partial S} &:= \Omega'_{S}\setminus\Omega_{S},
\end{aligned}
\end{equation}
where $\text{dilation}(\cdot)$ is the dilation function with a structuring element, which consists of a circle with a 20 pixel radius.
The overline $\overline{\cdot}$ is the complement and the operator $\setminus$ is the set difference.
An example of each region can be seen in Fig.~\ref{fig:eachregi}.

%

%
%

\subsection{Color grading by \cite{Pitie2007}}
\label{sec:grading}

\begin{figure}[t]
\centering
\setlength{\tabcolsep}{1mm}
\begin{tabular}{c}
\begin{tabular}{ccc}
\includegraphics[trim=30 30 30 30,width=.22\linewidth,clip]{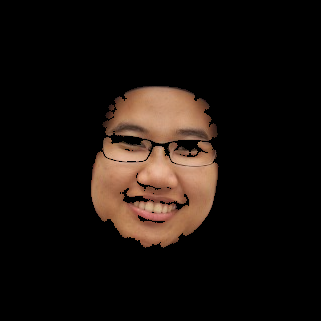}&
\includegraphics[trim=30 30 30 30,width=.22\linewidth,clip]{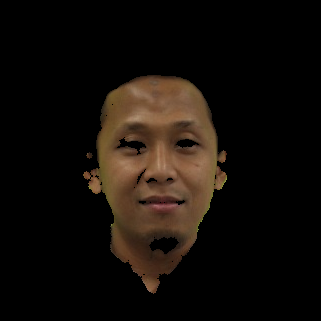}&
\includegraphics[trim=30 30 30 30,width=.22\linewidth,clip]{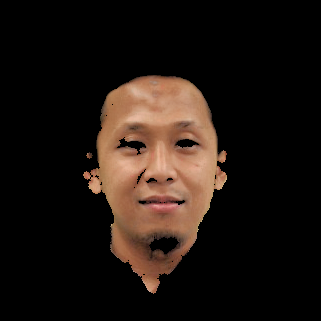}
\\
\includegraphics[trim=60 10 60 50,width=.31\linewidth,clip]{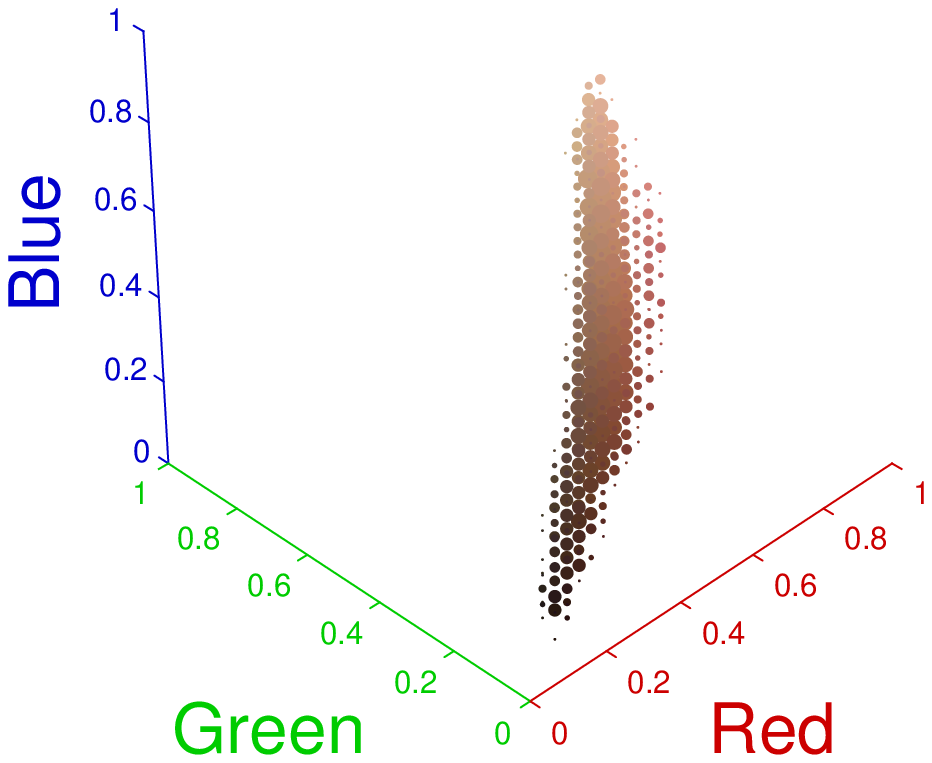}&
\includegraphics[trim=60 10 60 50,width=.31\linewidth,clip]{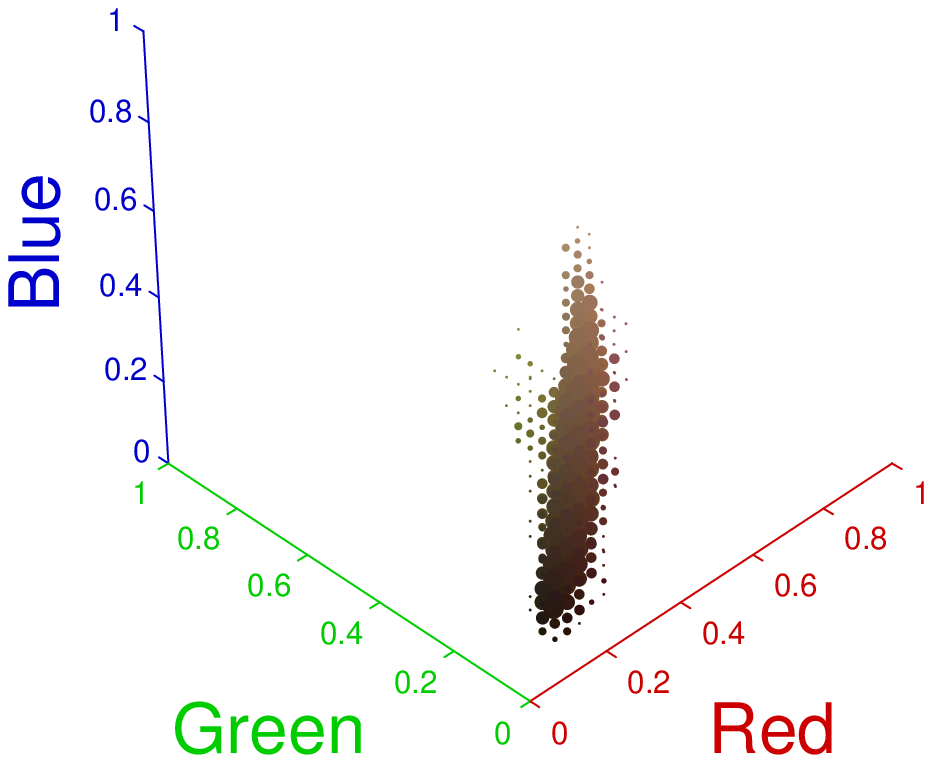}&
\includegraphics[trim=60 10 60 50,width=.31\linewidth,clip]{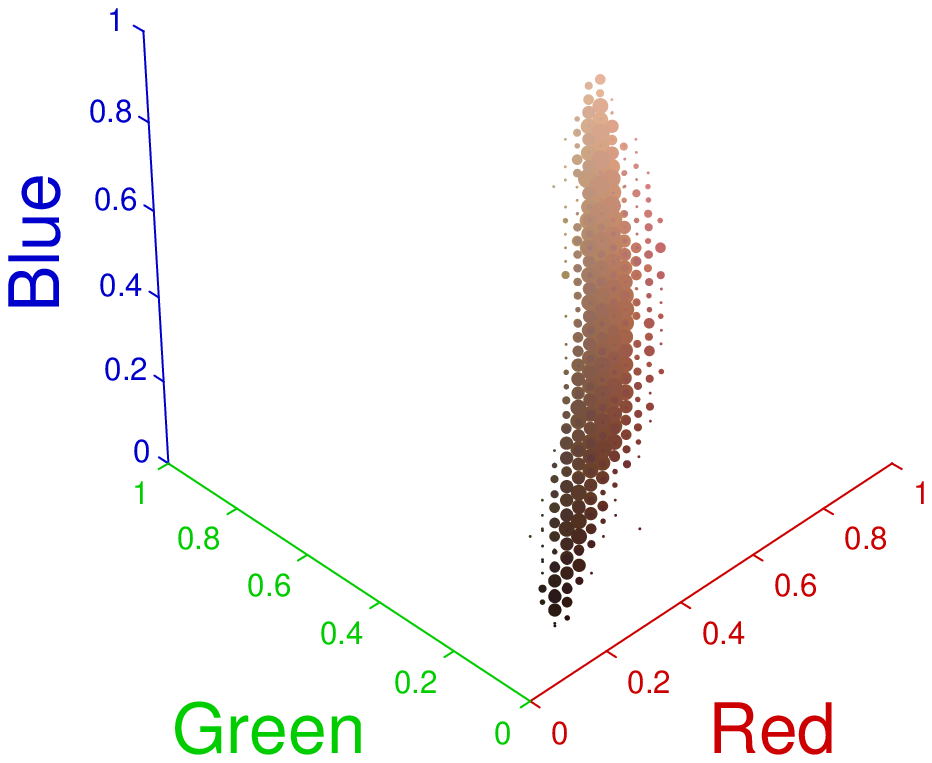}
\\
(a) Target & (b) Input & (c) Corrected
\end{tabular}
\end{tabular}
\caption{Each extracted facial skin color region and their distributions. The top row shows (left) target color image, (center) input image, and (right) color graded image. The bottom row shows their color distributions in the RGB color space.}
\label{fig:color_grade}
\end{figure}

Color grading (Fig.~\ref{fig:color_grade}) is performed so as to bring the skin color of the extracted region $\Omega_{S}$ (more specifically the shape of the color distribution\footnote{Note that we use the RGB color space instead of the HSV color space used in the preceding section. The reason is that the color grading method [1] has been developed for the RGB color space and the color-line image feature used in the next section is a feature related to the RGB color space.}) close to the target facial skin color in Fig.~\ref{fig:color_grade}(a). Color grading transforms the shapes of color distributions (r,g,b 3D coordinates).
One may use simpler techniques such as the estimation of a tone curve to modify each RGB component. However, most of the methods need correspondence between the two images, which is not suitable in our case where one transfers the color between the faces of different persons. In our method we use Piti{\'e} {\it et al.}'s method \cite{Pitie2007}, in which pixel correspondences are not necessary.
This method considers the width of the color distribution along a particular axis, iteratively changes it so as to match the specified axis, and then realizes nonlinear reshaping of the color distribution.
Since the method yields some artifacts such as blurring artifacts around the edges, we need to address this problem, which is described hereafter.

%

%
%

\subsection{Guide image filtering via optimization} \label{sec:guide}

This section expresses our GIF, which uses a guide image to correct the input image, as an optimization based method.
Our energy function actually resembles in part that of the image matting method \cite{Levin2008} because we focus on and use its data fidelity term (in \cite{Levin2008} that is also known as the \textit{local linear model}), but the design objective is different. Additionally, a similar method has been proposed in high dynamic range imaging \cite{Shan2010}.

\begin{figure}[t]
\centering
{\small
\setlength{\tabcolsep}{.5mm}
\begin{tabular}{ccccc}
\includegraphics[height=21mm]{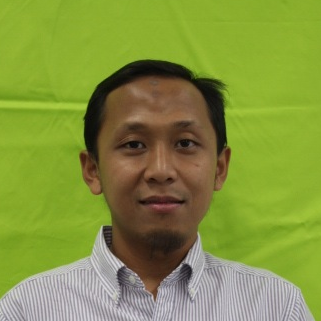}&
\includegraphics[height=21mm]{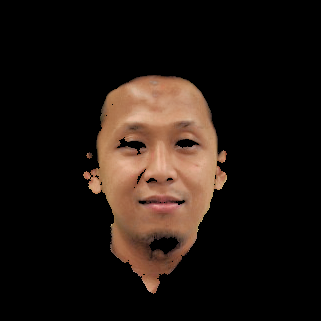}&
\includegraphics[height=21mm]{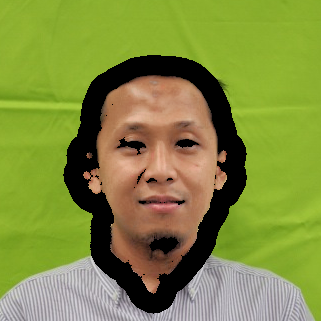}&
\includegraphics[height=21mm]{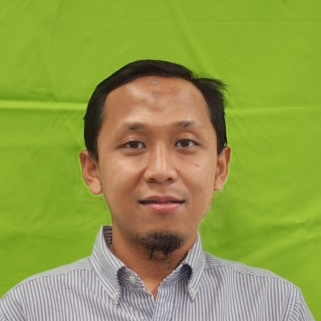}
\\
$\vec{y}$ & $\vec{g}_p,\ \forall p\in\Omega_{S}$ & \begin{tabular}{l}$\vec{g}_p,\ \forall p\in\Omega_{S}$\\$\vec{y}_q,\ \forall q\in\Omega_{B}$\end{tabular}  & $\vec{x}$\\
\end{tabular}
\caption{Guide image filtering. From left: Input image, Guide image, Target color in $\Omega_{B}$ and $\Omega_{S}$, and Color transformed image.}
\label{fig:guide_filter}
}
\end{figure}

Our guide image filtering reconstructs an image $\vec{x}\in\mathbb{R}^{3N}$ where the facial skin color region has a corrected color by using the input image $\vec{y}\in\mathbb{R}^{3N}$ and the color grading result, where $N$ is a number of pixels.
In the RGB color space ($\mathbb{R}^3$), let the pixel value at pixel $j$ of the color corrected image to be solved be $\vec{x}_j$, the input image be $\vec{y}_j$, and the guide image be $\vec{g}_j$ which are given by the color grading in Sec.~\ref{sec:grading}.
Figure~\ref{fig:guide_filter} shows each image. 
Using them, we formulate our guide image filtering as the following convex optimization problem
\begin{equation}
\begin{aligned}
\displaystyle \min_{\vec{x,A,b}}\ \underbrace{\sum_i \left( \sum_{j \in w_i}\|  \vec{x}_j - \vec{A}_i \vec{y}_j - \vec{b}_i  \|^2_2 + \varepsilon\|\vec{A}_i\|^2_F \right)}_{\text{data fidelity}}
\\ \ +\ \underbrace{\iota_{\mathcal{C}_{S}}( \vec{x} ) +\ \iota_{\mathcal{C}_{B}}( \vec{x} )}_{\text{constraints}},
\end{aligned}
\label{eq:llm}
\end{equation}
where $\vec{A}_i\in\mathbb{R}^{3\times 3}$ and $\vec{b}_i\in\mathbb{R}^3$ are a scaling matrix and an offsetting vector to approximate $\vec{y}_j$ to $\vec{x}_j$ for each pixel $i$, $w_i$ is a square window around pixel $i$, and $\mathcal{C}_{S}$ and $\mathcal{C}_{B}$ are given by
\begin{equation}
\begin{aligned}
\mathcal{C}_{S} &:= \textstyle \left\{ \vec{x} \mid \sqrt{ \sum_{p\in\Omega_{S}} \|\vec{x}_p-\vec{g}_p\|^2 } \leq \eta_{S} \right\},\\
\mathcal{C}_{B} &:= \textstyle \left\{ \vec{x} \mid \sqrt{ \sum_{q\in\Omega_{B}} \|\vec{x}_q-\vec{y}_q\|^2 } \leq \eta_{B} \right\}.
\end{aligned}
\end{equation}
Terms in the top row correspond to \cite{Levin2008}, where $\|\cdot\|_2$ and $\|\cdot\|_F$ denote the $\ell_2$ norm and Frobenius norm respectively, and we use them as a data fidelity that reflects textures and local contrasts of $\vec{y}$ onto $\vec{x}$.
The second and third terms are constraints expressed by the following indicator function:
\begin{equation}
\iota_{\mathcal{C}}(\vec{x}) :=
\left \{
\begin{array}{ll}
0 & \text{if} \ \vec{x} \in {\mathcal{C}}, \\
+\infty  & \text{otherwise}.
\end{array}
\right.
\end{equation}
%
The second term brings the facial skin color close to that of the guide image $\vec{g}$ in the facial skin color region $\Omega_{S}$.
The third term keeps the color of the background the same as the original image $\vec{y}$ in $\Omega_{B}$.
For reducing undesirable artifacts arising from the guide image, we add a constraint that the color difference at each pixel does not exceed $\eta$.

Note that we purposefully represent the second and third terms as constraint formulation over unconstrained one, \textit{e.g.}, the second term can be replaced with a regularizer $\lambda \sqrt{ \sum_{p\in\Omega_{S}} \|\vec{x}_p-\vec{g}_p\|^2 }$ where $\lambda$ is a balancing parameter. This is because $\eta$ can be controlled more intuitively than $\lambda$, and it allows us to adaptively change $\eta$ depending on the area of $\Omega_s$ (see the next Sec.~\ref{sec:results}).
Such advantages of the formulation have been addressed in the literature of image restoration based on convex optimization \cite{Combettes2004,Fadili2011,Afonso2011,Teuber2013,Ono2014_TGV,Chierchia2015,Ono2015a}.

Among convex optimization algorithms, we adopt \textit{a monotone version of the fast iterative shrinkage-thresholding algorithm} (MFISTA) \cite{mfista} to solve \eqref{eq:llm} because it is a first-order method that achieves a sublinear global rate of convergence, namely, computationally efficient and very fast.
The algorithm solves an optimization problem of the form:
\begin{equation}
\underset{\vec{x}}{\min}\ f(\vec{x}) + h(\vec{x}),
\label{eq:mfista}
\end{equation}
where $f(\vec{x})$ is a differentiable convex function with a Lipschitz continuous gradient and $h(\vec{x})$ is a proper lower semicontinuous convex function.
The problem \eqref{eq:mfista} is calculated by the proximity operator
\footnote{
The proximity operator function is defined by
\[\text{prox}_{\kappa h}(y) := \arg\underset{x}{\min}\  \kappa h (x) + \frac{1}{2}\| y-x \|_2^2. \]
where $h(\cdot)$ is a proper lower semicontinuous convex function and $\kappa$ is the index.
}.
For given $\vec{x}^0=:\vec{v}^1\in \mathbb{R}^N$ and $t^1:=1$, the iteration of the MFISTA consists of the following five steps:
%
\begin{equation}
\left\lfloor
\begin{aligned}
\hat{\vec{z}}^k &= \displaystyle  \vec{v}^k - \frac{1}{L}\nabla f(\vec{v}^k),
\\
\vec{z}^k &= \displaystyle \text{prox}_{h/L}\left( \hat{\vec{z}}^k \right),
\\
\vec{x}^k &= \displaystyle \arg\min\{\ f(\vec{z}) + h(\vec{z})\ |\ \vec{z}\in\{ \vec{z}^k,\vec{x}^{k-1}\}\},
\\
t^{k+1} &= \displaystyle \frac{1+\sqrt{1+4(t^k)^2}}{2},
\\
\vec{v}^{k+1} &= \displaystyle \vec{x}^k + \frac{t^k}{t^{k+1}}(\vec{z}^k-\vec{x}^k)+\frac{t^k-1}{t^{k+1}}(\vec{x}^k-\vec{x}^{k-1}),
\end{aligned}
\right.
\label{eq:mfistastep}
\end{equation}
where ``$\text{prox}$'' is the proximity operator and $1/L$ is the step size.

In order to apply the MFISTA to our problem \eqref{eq:llm}, we set
\begin{equation}
\begin{aligned}
f(\vec{x}) &:= \displaystyle\sum_i \left( \sum_{j \in w_i}\|  \vec{x}_j - \vec{A}_i \vec{y}_j - \vec{b}_i  \|^2_2 + \varepsilon\|\vec{A}_i\|^2_F \right),\\
h(\vec{x})&:= \iota_{\mathcal{C}_{S}}( \vec{x} ) +\ \iota_{\mathcal{C}_{B}}( \vec{x} ).
\end{aligned}
\end{equation}
To compute the gradient $\nabla f(\vec{x})$, we use a method similar to \cite{He2010}, which is an accelerated version of \cite{Levin2008}.
We compute each value of $[\nabla f(\vec{x})]_i\in \mathbb{R}^{3}$ as follows:
\begin{align}
\vec{A}^\ast_i  & = \Delta_i^{-1}\left(\frac{1}{|w_i|}\left(\displaystyle \sum_{j\in w_i} \vec{y}_j\vec{x}_j^\top \right) - \overline{\vec{y}}_i\overline{\vec{x}}_i^\top \right),
\\
\vec{b}^\ast_i & = \overline{\vec{x}}_i - \vec{A}_i^\ast\overline{\vec{y}}_i,
\\
[\nabla f(\vec{x})]_i & = |w_i|\vec{y}_i - \left(\left(\sum_{j\in w_i}\vec{A}_j^\ast\right) \vec{y}_i + \left(\sum_{j\in w_i}\vec{b}_j^\ast\right)\right),
\end{align}
where $\overline{\vec{x}}_i = [\overline{x}^R_i\ \overline{x}^G_i\ \overline{x}^B_i ]^\top$ and $\overline{\vec{y}}_i = [\overline{y}^R_i\ \overline{y}^G_i\ \overline{y}^B_i ]^\top$ are mean value vectors of each color in a square window $w_i$, $|w_i|$ is the number of pixels in $w_i$, $\Delta = \bsym{\Sigma}_i+\frac{\varepsilon}{|w_i|}\vec{U} \in\mathbb{R}^{3\times 3}$, $\bsym{\Sigma}_i = \frac{1}{|w_i|}\left( \sum_{j\in w_i} \vec{y}_j\vec{y}_j^\top \right) - \overline{\vec{y}}_i\overline{\vec{y}}_i^\top $ is a covariance matrix, and $\vec{U}$ is an identity matrix.

The proximity operator $\text{prox}_{h/L}$ in \eqref{eq:mfistastep} consists of two functions which are the second and third terms of \eqref{eq:llm} which handle different regions, $\Omega_{S}$ and $\Omega_{B}$, respectively that satisfies, \textit{i.e.}, $\Omega_{S}\cap\Omega_{B}=\{ 0 \}$. Therefore, $\text{prox}_{h/L}$ can be calculated using
\begin{equation}
\begin{aligned}
& \vec{z}_i:=[\text{prox}_{h/L}(\hat{\vec{z}})]_i=
\\
& \begin{cases}
\ \displaystyle \hat{\vec{z}}_i + \eta_{S}\frac{\hat{\vec{z}}_i-\vec{g}_i}{\sqrt{ \sum_{p\in\Omega_{S}} \|\hat{\vec{z}}_p-\vec{g}_p\|^2 }} &
\text{if}\ \hat{\vec{z}} \notin \mathcal{C}_{S} \land\ i \in \Omega_{S},
\\
\ \displaystyle \hat{\vec{z}}_i + \eta_{B}\frac{\hat{\vec{z}}_i-\vec{y}_i}{\sqrt{ \sum_{q\in\Omega_{B}} \|\hat{\vec{z}}_q-\vec{y}_q\|^2 }} &
\text{if}\ \hat{\vec{z}} \notin \mathcal{C}_{B} \land\ i \in \Omega_{B},
\\
\ \hat{\vec{z}}_i &\text{otherwise}.
\end{cases}
\end{aligned}
\end{equation}
%
This process corresponds to a $\ell_2$-ball projection with a region constraint.

%
Finally, we update $\vec{x, z, v}, t$ using the procedure in \eqref{eq:mfistastep}, then the solution $\vec{x}$ becomes our GIF result.

%
%
%

\section{Results and discussion} \label{sec:results}

In this section, we show the results obtained through the proposed process
\footnote{
The experiment tests our algorithm on various sets, which are available from our website http:vig.is.env.kitakyu-u.ac.jp/MIF/.
}.
The range of RGB values is normalized to be in the range $[0,1]$. The prescribed scale factor $l$ in Sec.~\ref{sec:face_detect} is set to $2.0$. The filter window sizes used in Sec.~\ref{sec:guide} and Sec.~\ref{sec:matting} are $19 \times 19$ and $31 \times 31$, respectively.
$\eta_{S}=5|\Omega_{S}|\times 10^{-4}$ and $\eta_{B}=5|\Omega_{B}|\times 10^{-10}$ are used, where $|\Omega_{(\cdot)}|$ indicates the number of pixels contained in the region $\Omega_{(\cdot)}$.
$L$ in MFISTA is set to 500.

\begin{figure}[t]
\centering
\includegraphics[width=\linewidth]{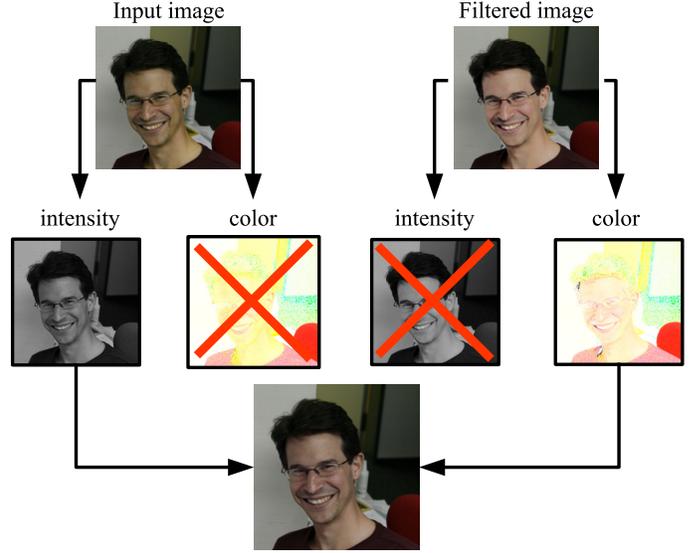} 
\caption{Simple luminance correction. Input image and filtered image are decomposed into intensity components and color components, respectively by \eqref{eq:decompose}, then the intensity component of the input image and the color component of the filtered image are combined by \eqref{eq:combine}.}
\label{fig:shadowtreatment}
\end{figure}
\begin{figure*}[p]
\centering
{\small
\setlength{\tabcolsep}{1mm}
\begin{tabular}{cc}
\begin{tabular}{c}
\includegraphics[width=21mm]{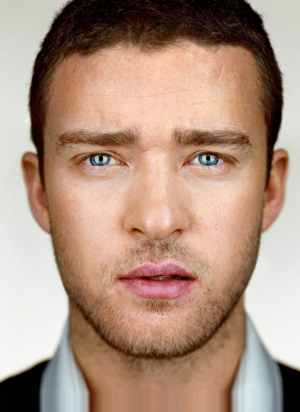}
\\
(a) Target 
\end{tabular}
&
\begin{tabular}{ccc}
\includegraphics[trim=0 25 0 25,width=0.27\linewidth,clip]{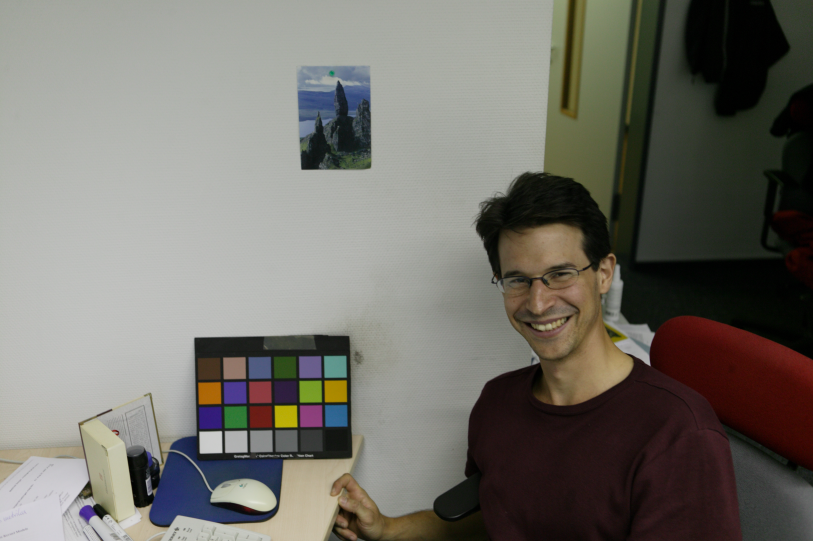}&
\includegraphics[trim=0 25 0 25,width=0.27\linewidth,clip]{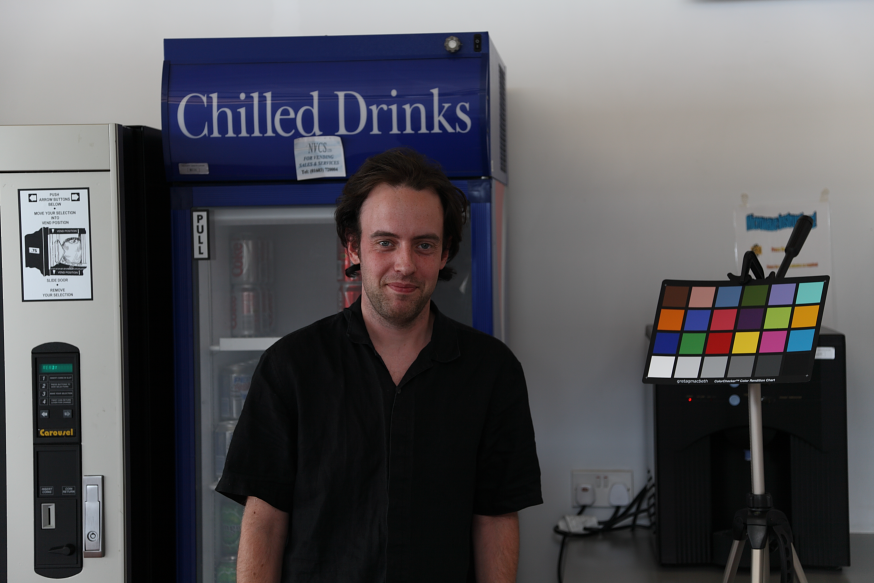}&
\includegraphics[trim=0 25 0 25,width=0.27\linewidth,clip]{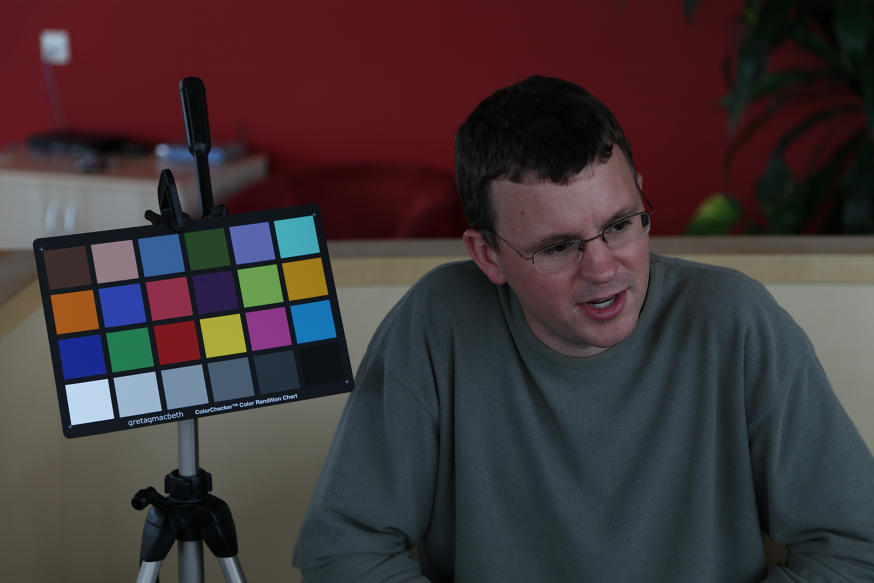} 
\\
\multicolumn{3}{c}{(b) Input}
\\
\includegraphics[trim=0 25 0 25,width=0.27\linewidth,clip]{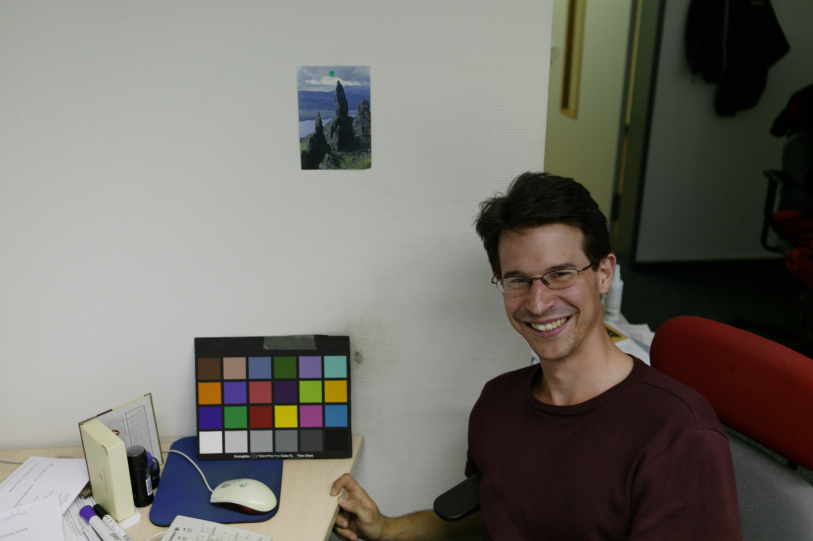}&
\includegraphics[trim=0 25 0 25,width=0.27\linewidth,clip]{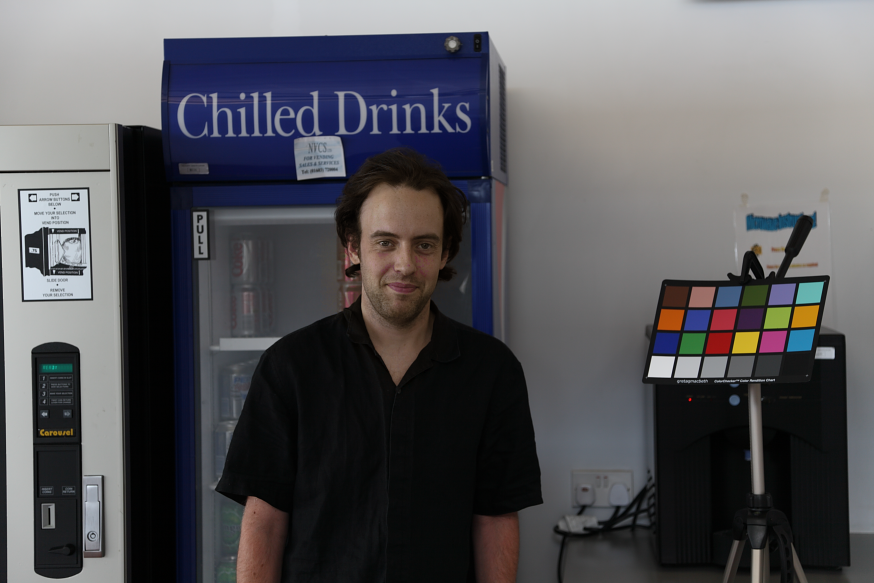}&
\includegraphics[trim=0 25 0 25,width=0.27\linewidth,clip]{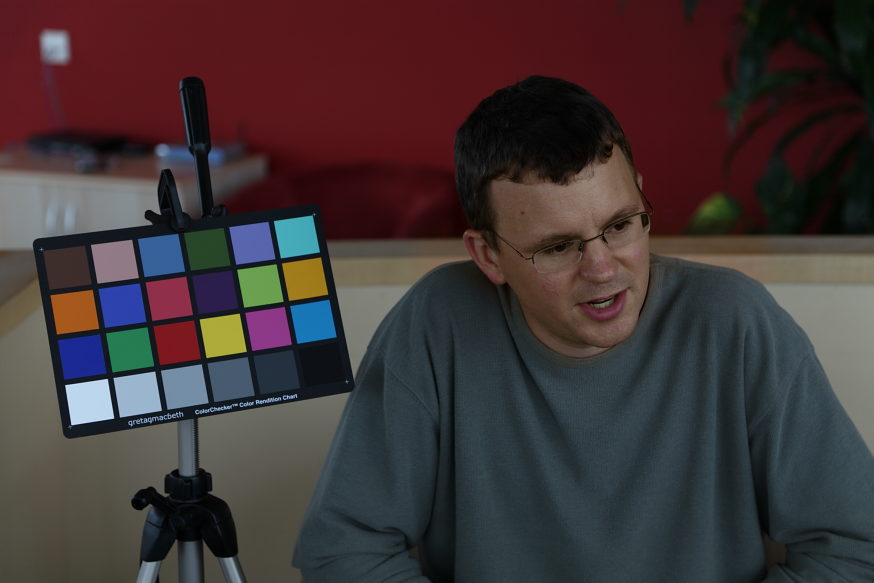}
\\
\multicolumn{3}{c}{(c) Output}
\end{tabular}
\end{tabular}
}
\caption{Our facial skin color correction. (a) Target facial skin color image, (b) Input image, and (c) Output image using our facial skin color correction. The target image is in Shin \textit{et al.}'s dataset \cite{Shin2014} and the input images are in the Gehler dataset \cite{Gehler}.}
\label{fig:daily}
%
\vspace{2\baselineskip}
%
\centering
{\small
\setlength{\tabcolsep}{1mm}
\begin{tabular}{cc}
\begin{tabular}{c}
\includegraphics[trim=20 0 20 0,width=21mm,clip]{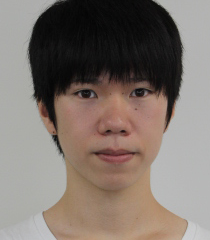}
\\
(a) Target
\end{tabular}
&
\begin{tabular}{ccc}
\includegraphics[trim=0 25 0 25,width=0.27\linewidth,clip]{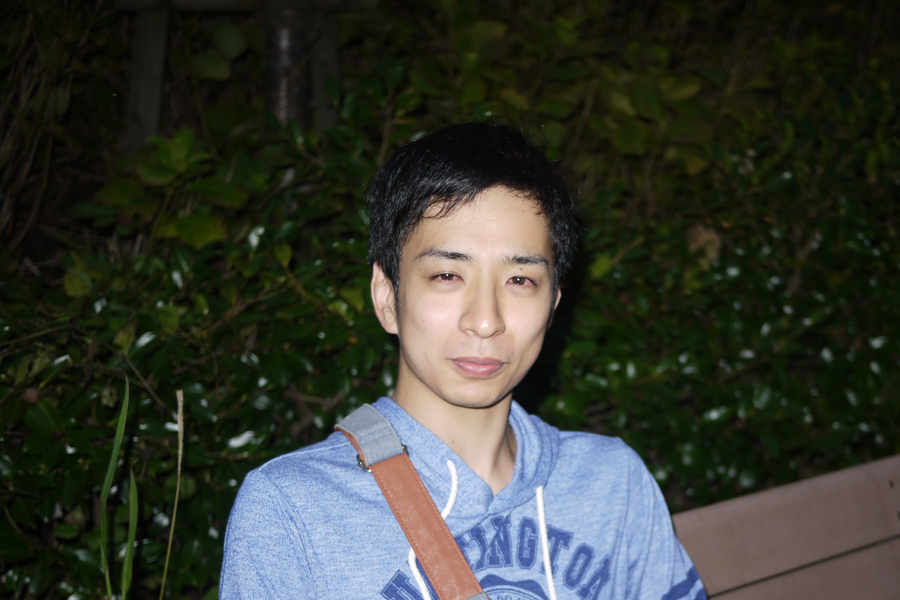}&
\includegraphics[trim=0 25 0 25,width=0.27\linewidth,clip]{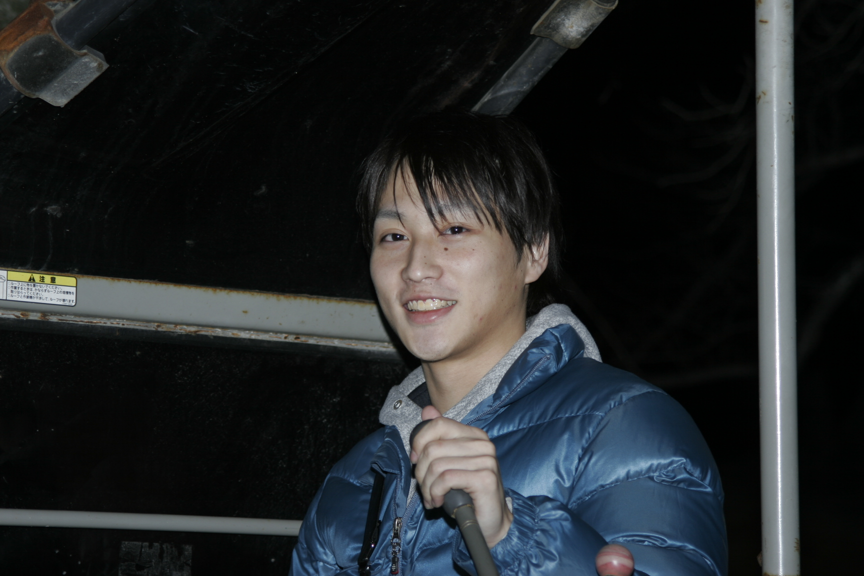}&
\includegraphics[trim=300 100 100 255,width=0.27\linewidth,clip]{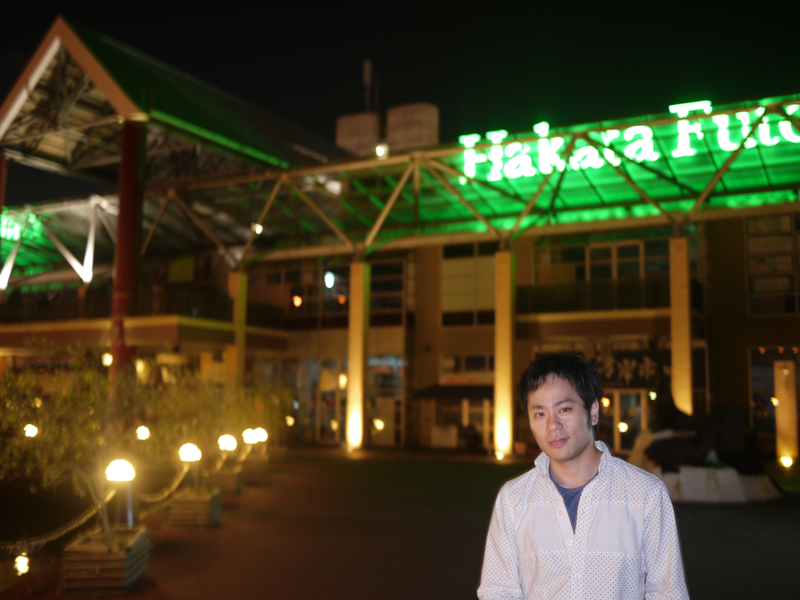} 
\\
\multicolumn{3}{c}{(b) Input}
\\
\includegraphics[trim=0 25 0 25,width=0.27\linewidth,clip]{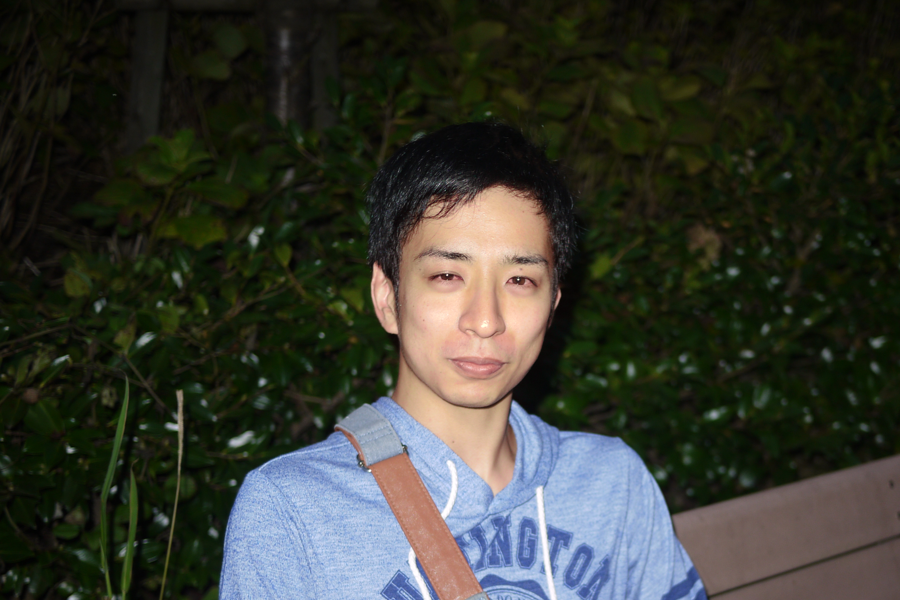}&
\includegraphics[trim=0 25 0 25,width=0.27\linewidth,clip]{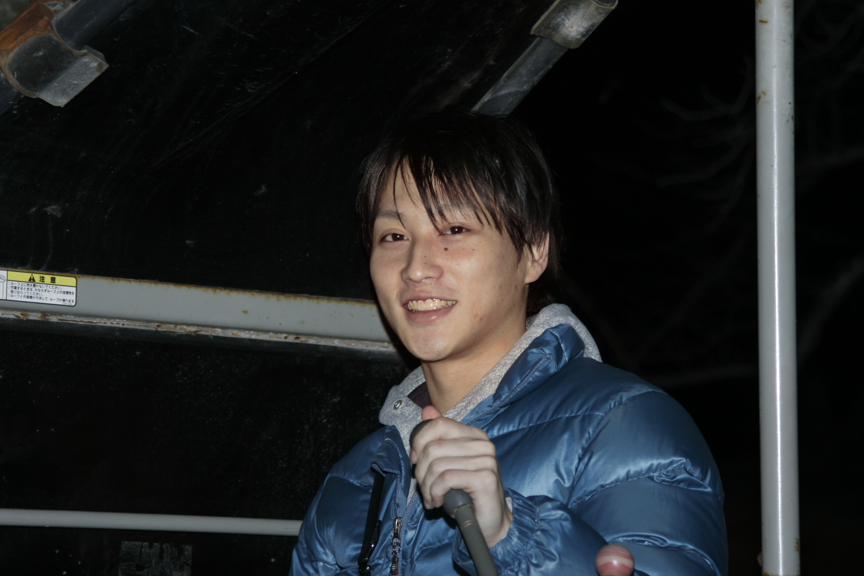}&
\includegraphics[trim=300 100 100 255,width=0.27\linewidth,clip]{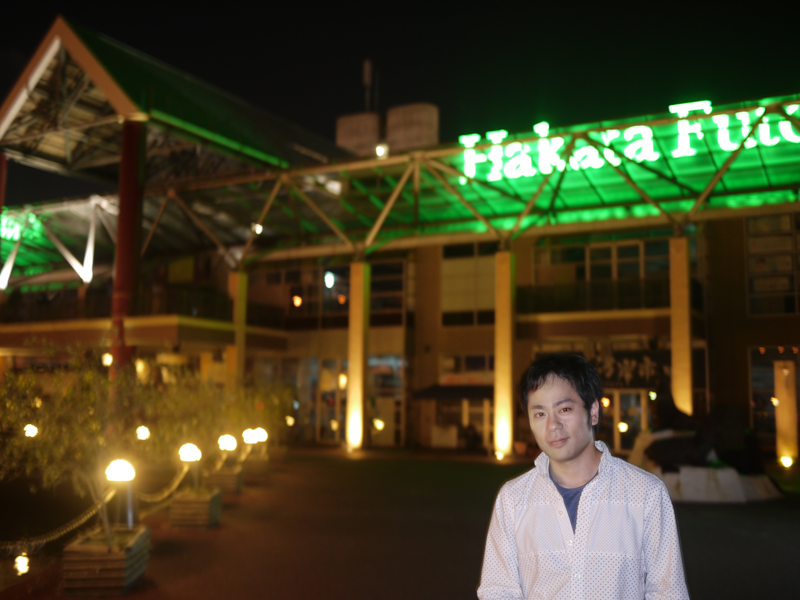}
\\
\multicolumn{3}{c}{(c) Output} 
\end{tabular}
\end{tabular}
}
\caption{Our facial skin color correction on the flash images. (a) Target facial skin color image, (b) Input flash images, (c) Output images using our facial skin color correction.}
\label{fig:flash}
\end{figure*}

Our filtering often flattens the gradation of the input images caused by shadows. It yields unnatural images as shown in the top of Fig.~\ref{fig:shadowtreatment}, where the input image is input to our algorithm and the filtered image is a result of our guide image filtering.
For the luminance correction, each pixel color value of the image is decomposed into a color component $\vec{x}_i^{C}\in\mathbb{R}^3$ and an intensity component $x_i^I\in\mathbb{R}^1$ as follows
\begin{equation}
\begin{aligned}
x_i^I &=\displaystyle  \frac{x_i^R}{\|\vec{x}_i\|_1}x_i^R + \frac{x_i^G}{\|\vec{x}_i\|_1}x_i^G + \frac{x_i^B}{\|\vec{x}_i\|_1}x_i^B,
\\
\vec{x}_i^{C} & = \displaystyle  \frac{\vec{x}_i}{x_i^I},
\end{aligned}
\label{eq:decompose}
\end{equation}
where the superscripts $R$,$G$, and $B$ indicate each color component.
This decomposition procedure is same as \cite{Eisemann2004}.

The input image and the filtered image are decomposed into the two components by \eqref{eq:decompose}, and then the intensity component of the input image $y_i^{I}$ and the color component of the filtered image $\vec{x}_i^{C}$ are combined as follows
\begin{equation}
{\vec{x}_i}' = y_i^{I}\vec{x}_i^{C}.
\label{eq:combine}
\end{equation}
Figure~\ref{fig:shadowtreatment} shows this procedure and its effectiveness.

Figure~\ref{fig:daily} (hereafter indexes are swapped due to figure positions) shows the result of our skin color correction. The area around the face in the input image has color distortion due to the lighting conditions and background color. Our result has a white balanced facial skin color similar to the target facial skin color. 
Sometimes when we take a photograph in dark surroundings, the results sometimes have unnatural face colors (Fig.~\ref{fig:flash}(b)) due to the camera flash.
Hence, we apply our proposed method to flash images in dark surroundings to reduce undesirable effects of artificial lights.
Figure~\ref{fig:flash}(c) shows the flash image editing results using our method.
One can see that unnatural colors of the original image are corrected to natural colors by our method.

%
%

\subsection{Automatic yearbook style photo generation}\label{sec:yearbook}

This section presents automatic yearbook style photo generation using our guided facial skin correction method and pre/post-processing procedure.
It takes a long time to manually process a large number of images.
Our algorithm generates a yearbook style photo in a short amount of time.

\begin{figure}[t]
\centering
\includegraphics[width=\linewidth]{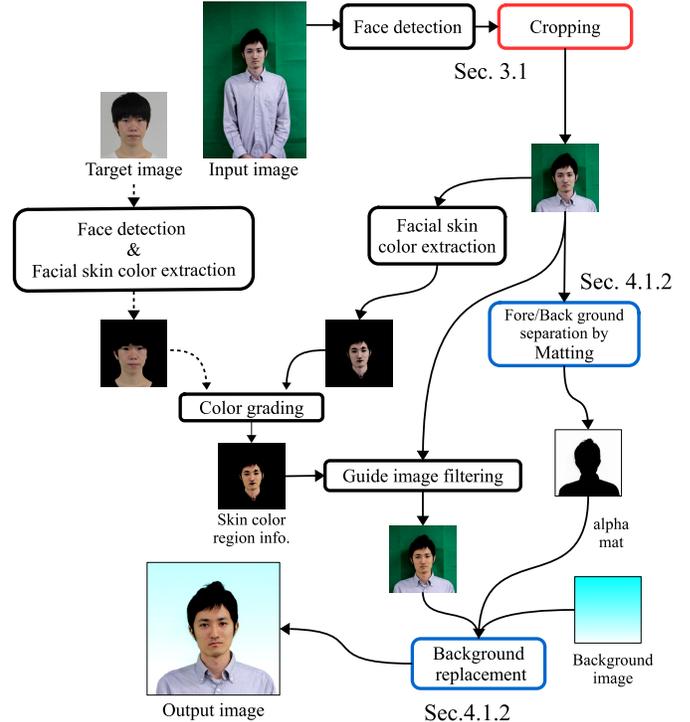} 
\caption{Flow chart of auto yearbook style photo generation. The red box indicates a pre-processing procedure and the blue boxes indicate a post-processing procedure.}
\label{fig:yearbookflow}
\end{figure}

We first crop a photo with our face detection procedure as a pre-processing procedure (the red box in Fig.~\ref{fig:yearbookflow}), and then correct the facial skin color. Finally, the noisy background is replaced to clear the background as a post-processing procedure (the blue boxes in Fig.~\ref{fig:yearbookflow}). 

%
%

\subsubsection{Face area cropping}\label{sec:cropping}

In Sec.~\ref{sec:face_detect}, we detect the face area of a photo.
Then we unify the sizes of the cropped images by expansion and reduction.
The crop size is roughly adjusted according to the image size.
The image size after cropping and resizing is $320 \times 320$ ($h \times w$) in this experiment.

%
%

\subsubsection{Background replacement by alpha blending}\label{sec:bg_replace}

We separate the image information of the foreground and background regions and assign a value $\alpha_i:=[0,1]$ to them. 
Using the values as labels, the relationship between the foreground $\vec{f}_i$, background $\vec{b}_i$ (different from that in \eqref{eq:llm}), and original image $\vec{y}_i$ at each pixel is given as follows:
\begin{equation}
\vec{y}_i := \vec{f}_i + \vec{b}_i,
\quad
\vec{f}_i := (1-\alpha_i) \vec{y}_i,
\quad
\vec{b}_i := \alpha_i \vec{y}_i.
\end{equation}
The label $\alpha$ at each pixel is the blending rate, and called alpha-mat.
Replacement of the background with another background $\vec{z}$ is performed by
\begin{equation}
\vec{y}'_i := \vec{f}_i + \alpha_i \vec{z}_i.
\end{equation}
An estimation of the alpha-mats is described in Appendix A.

\begin{figure*}[p]
\centering
\setlength{\tabcolsep}{1mm}
\begin{tabular}{cc}
\begin{tabular}{c}
\includegraphics[width=33mm]{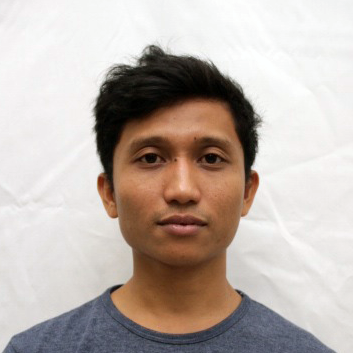}
\\
(a) Target
\end{tabular}
&
\setlength{\tabcolsep}{.3mm}
\begin{tabular}{cccc}
\includegraphics[width=33mm]{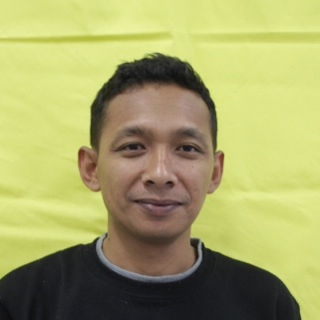}&
\includegraphics[width=33mm]{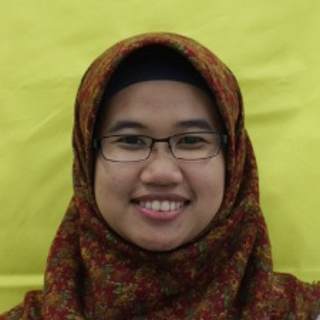}&
\includegraphics[width=33mm]{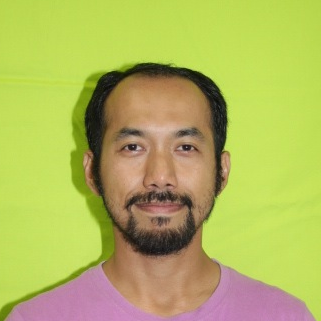}&
\includegraphics[width=33mm]{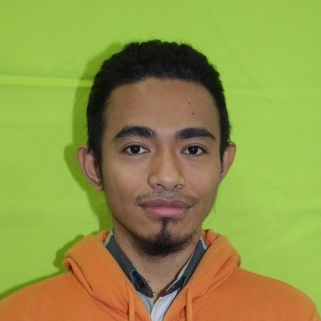}
\\
\multicolumn{4}{c}{(b) Original images}
\end{tabular}
\\
&
\setlength{\tabcolsep}{.3mm}
\begin{tabular}{cccc}
\includegraphics[width=33mm]{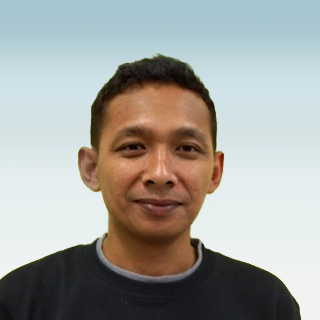}&
\includegraphics[width=33mm]{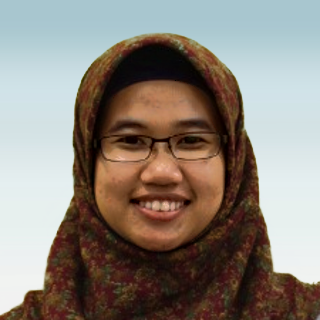}&
\includegraphics[width=33mm]{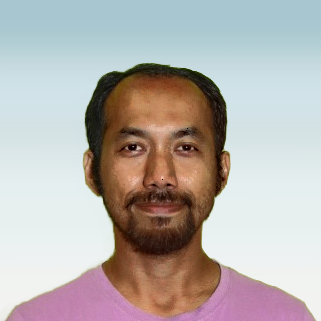}&
\includegraphics[width=33mm]{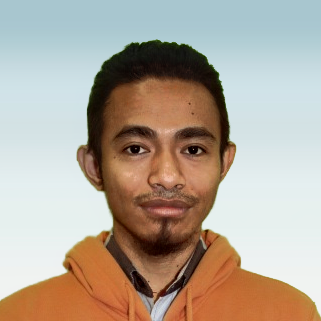}
\\
\multicolumn{4}{c}{(c) Yearbook style images}
\end{tabular}
\end{tabular}
\caption{Results of automatic yearbook style photo generation using our method. (a) Target facial skin color image, (b) Original images which have a background color similar to the facial skin color, (c) Yearbook style images using our algorithm.}
\label{fig:yearbook}
\end{figure*}
\begin{figure*}[p]
\centering
{\small
\setlength{\tabcolsep}{1mm}
\begin{tabular}{ccccc}
\includegraphics[width=33mm,clip]{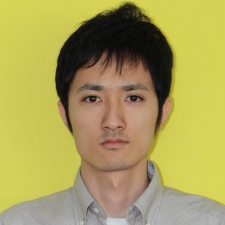} &
\includegraphics[width=33mm,clip]{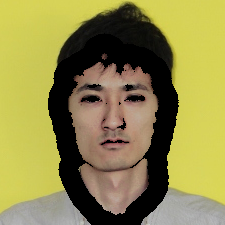} &
\includegraphics[width=33mm,clip]{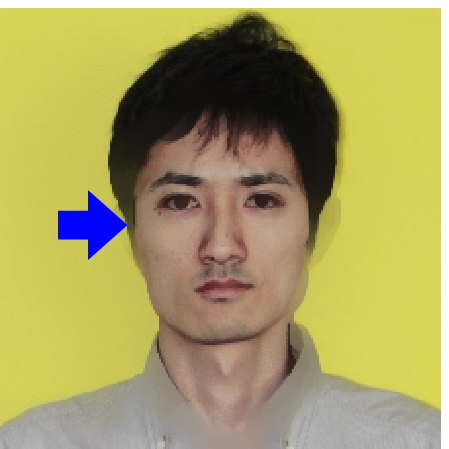} &
\includegraphics[width=33mm,clip]{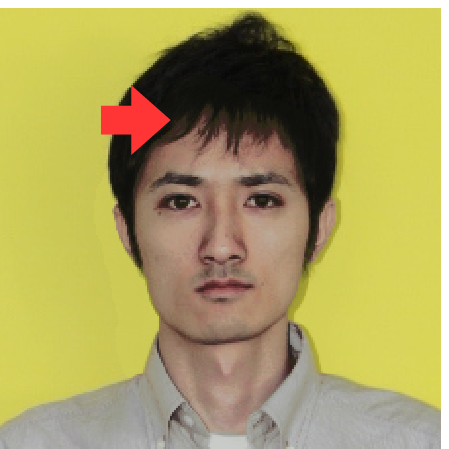} &
\includegraphics[width=33mm,clip]{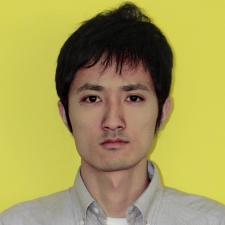} 
\\
\includegraphics[width=33mm,clip]{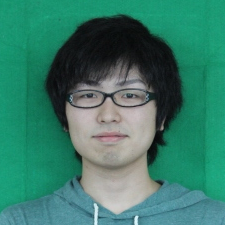} &
\includegraphics[width=33mm,clip]{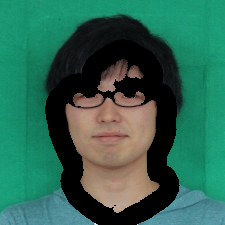} &
\includegraphics[width=33mm,clip]{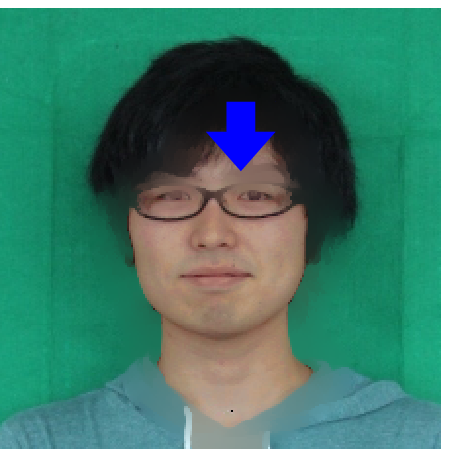} &
\includegraphics[width=33mm,clip]{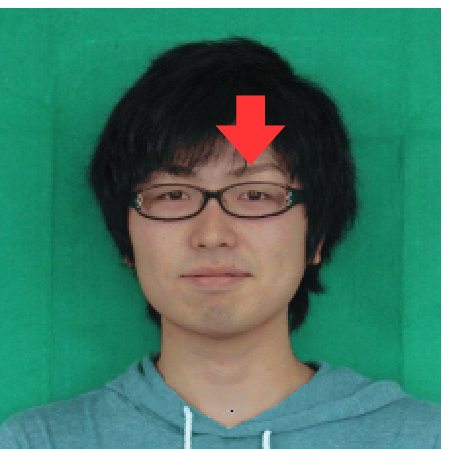} &
\includegraphics[width=33mm,clip]{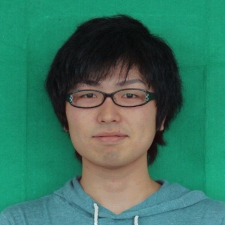} 
\\
\includegraphics[width=33mm,clip]{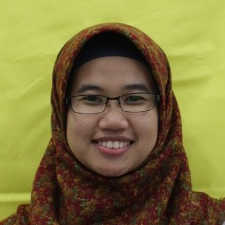} &
\includegraphics[width=33mm,clip]{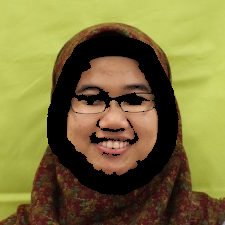} &
\includegraphics[width=33mm,clip]{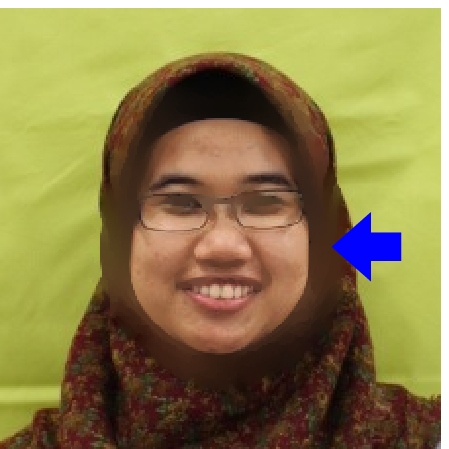} &
\includegraphics[width=33mm,clip]{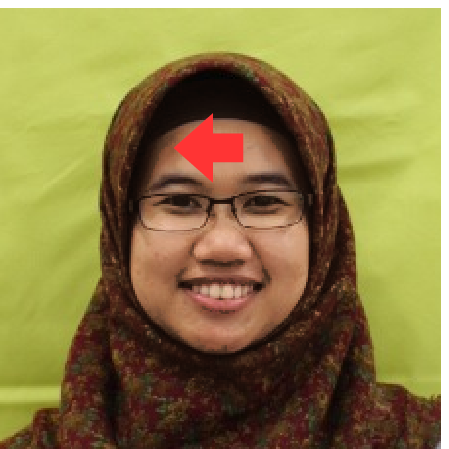} &
\includegraphics[width=33mm,clip]{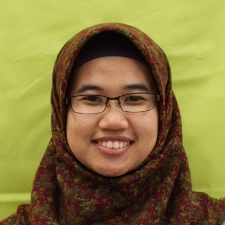} 
\\
(a) Original & (b) Guide & (c) Colorization \cite{Levin2004} & (d) JBU \cite{JBU} & (e) Ours
\end{tabular}
}
\caption{Supplement quality comparison with similar methods. (a) Original, (b) Guide, (c) Colorization for each RGB layer \cite{Levin2004}, (d) Joint Bilaterl Upsampling (JBU) for each RGB layer \cite{JBU}, and (e) Our guided filtering \eqref{eq:llm}. The black pixels in (b) represents a hole region and each red arrow indicates an artifact.}
\label{fig:result_compare1}
\end{figure*}

Figure~\ref{fig:yearbook} shows the results of our automatic yearbook style photo generation. 
Our algorithm generates the yearbook style photos from the original images using the target image.

We implement the whole algorithm in MATLAB and OpenCV (C++), respectively, and the total execution time is within 11 sec. on a 3.20 GHz Core i5 CPU, \textit{i.e.}, 
face detection (Sec.~\ref{sec:face_detect}) is 5 sec., facial color extraction (Sec.~\ref{sec:skin_extract}) is 1 sec., color grading (Sec.~\ref{sec:grading}) is 2 sec., GIF (Sec.~\ref{sec:guide}) is 2 sec., and matting (Sec.~\ref{sec:bg_replace}) is 1 sec., respectively.

%
%

\subsection{Comparison with various conventional method}
\label{sec:additional}

\begin{figure*}[p]
\centering
{\small
\setlength{\tabcolsep}{1mm}
\begin{tabular}{cccccc}
\includegraphics[trim=45 10 45 20,width=28mm,clip]{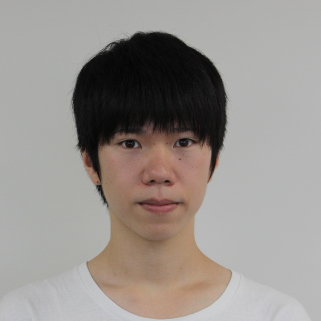} &
\includegraphics[trim=40 20 40 20,width=28mm,clip]{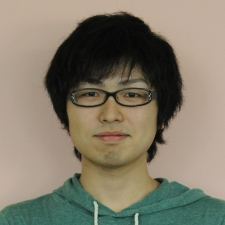} &
\includegraphics[trim=40 20 40 20,width=28mm,clip]{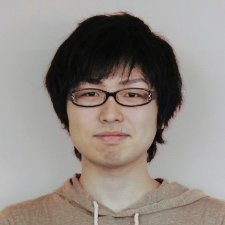} &
\includegraphics[trim=40 20 40 20,width=28mm,clip]{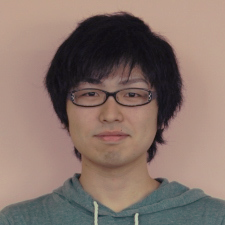} &
\includegraphics[trim=40 20 40 20,width=28mm,clip]{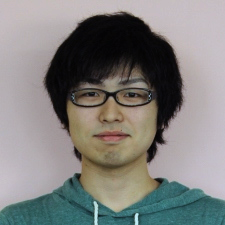} &
\includegraphics[trim=40 20 40 20,width=28mm,clip]{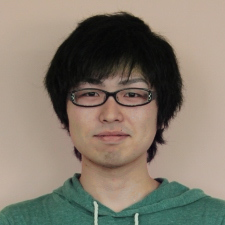} 
\\
\includegraphics[trim=70 0 70 50,width=28mm,clip]{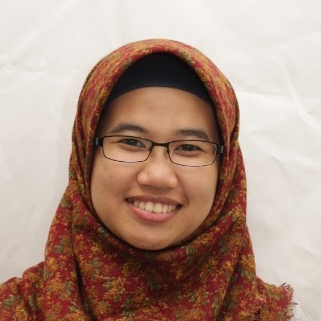} &
\includegraphics[trim=70 0 70 50,width=28mm,clip]{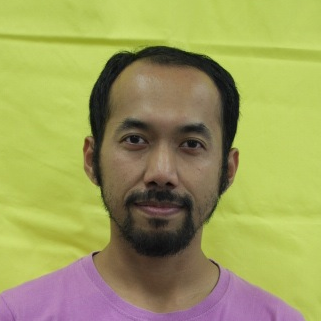} &
\includegraphics[trim=70 0 70 50,width=28mm,clip]{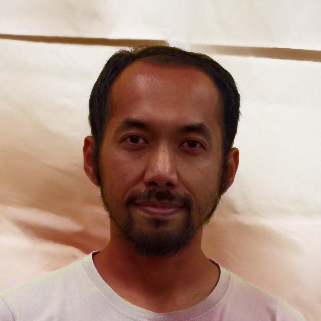} &
\includegraphics[trim=70 0 70 50,width=28mm,clip]{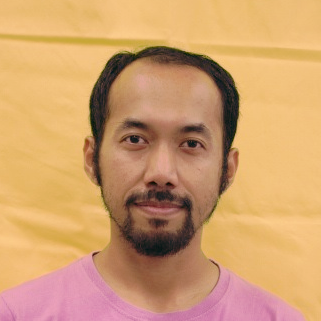} &
\includegraphics[trim=70 0 70 50,width=28mm,clip]{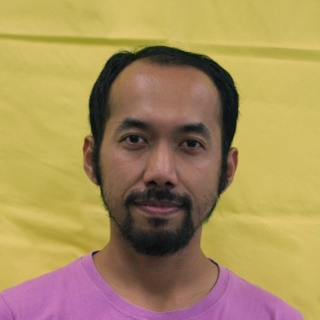} &
\includegraphics[trim=70 0 70 50,width=28mm,clip]{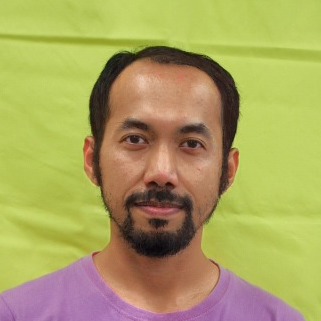} 
\\
\includegraphics[trim=40 0 40 0,width=28mm,clip]{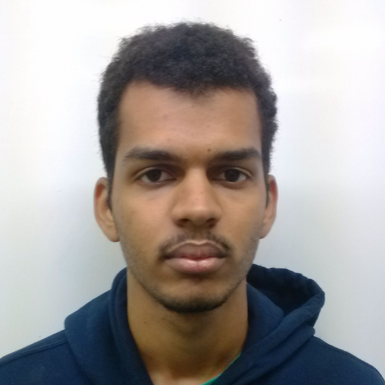} &
\includegraphics[trim=40 0 40 0,width=28mm,clip]{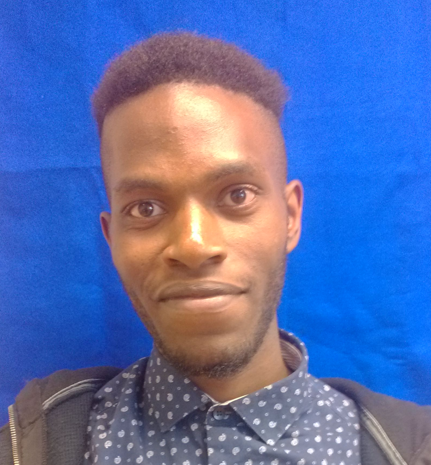} &
\includegraphics[trim=40 0 40 0,width=28mm,clip]{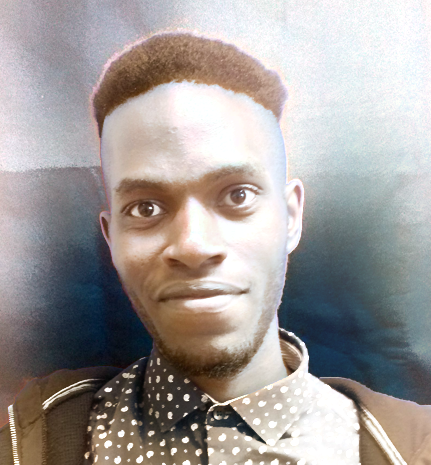} &
\includegraphics[trim=40 0 40 0,width=28mm,clip]{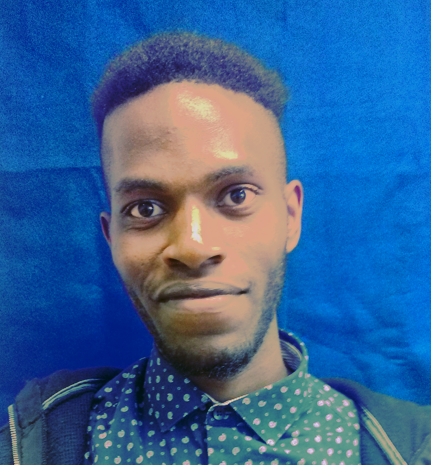} &
\includegraphics[trim=40 0 40 0,width=28mm,clip]{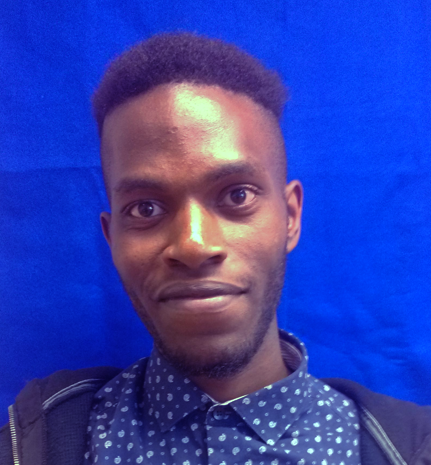} &
\includegraphics[trim=40 0 40 0,width=28mm,clip]{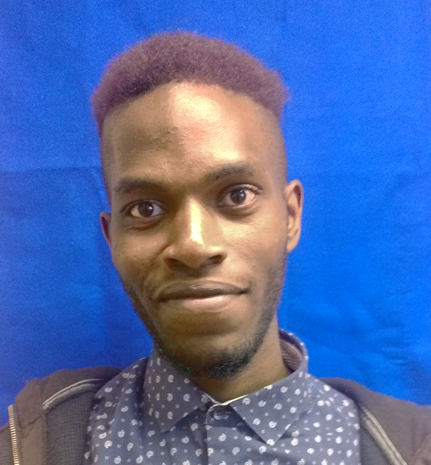} 
\\
(a) Target & (b) Original & (c) \cite{Rabin2010} with \cite{Pitie2007} & (d) \cite{HaCohen2011} & (e) \cite{Jaesik2016} & (f) Ours 
\end{tabular}
}
\caption{Comparison with the existing methods. (a) Target, (b) Original, (c) \cite{Rabin2010} with \cite{Pitie2007}, (d) NRDC \cite{HaCohen2011}, (e) Jaesik {\it et al.} \cite{Jaesik2016}, and (f) Our method. }
\label{fig:result_compare}
\end{figure*}
\begin{figure*}[p]
\centering
\setlength{\tabcolsep}{.5mm}
\begin{tabular}{ccc}
\begin{tabular}{c}
\includegraphics[width=19mm]{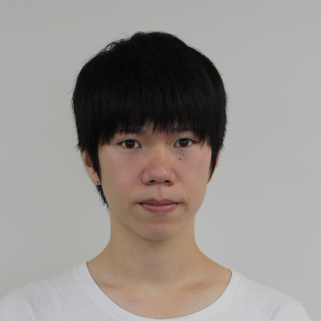}
\\
(a) Target
\end{tabular}
&
\setlength{\tabcolsep}{.3mm}
\begin{tabular}{cccc}
\includegraphics[width=19mm]{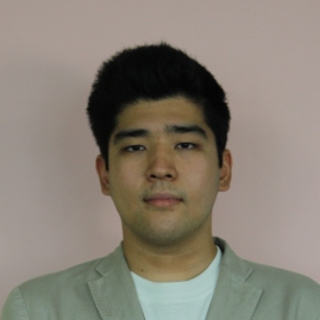}&
\includegraphics[width=19mm]{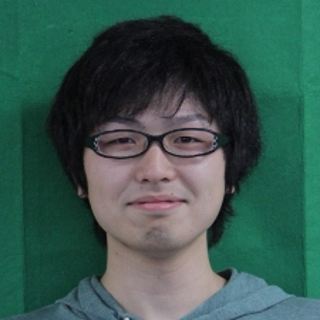}&
\includegraphics[width=19mm]{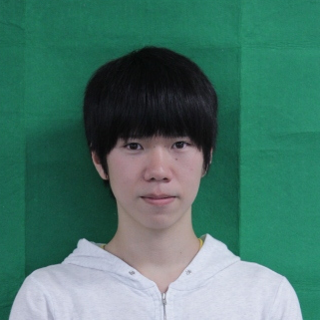}&
\includegraphics[width=19mm]{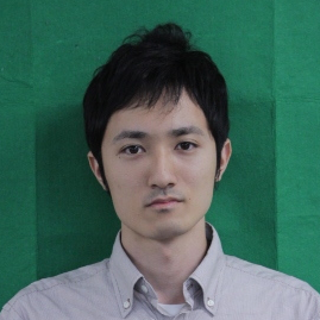}
\\
\multicolumn{4}{c}{(b) Original}
\end{tabular}
&
\setlength{\tabcolsep}{.3mm}
\begin{tabular}{cccc}
\includegraphics[width=19mm]{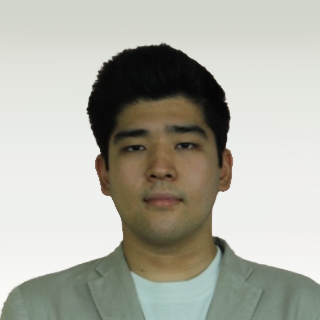}&
\includegraphics[width=19mm]{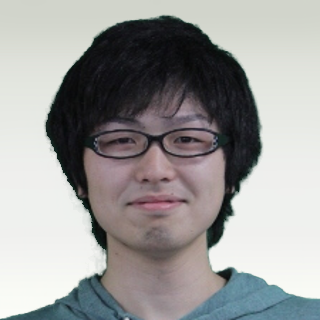}&
\includegraphics[width=19mm]{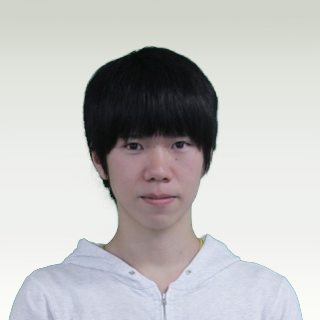}&
\includegraphics[width=19mm]{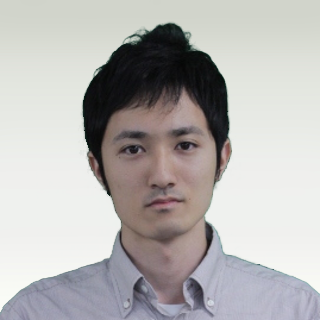}
\\
\multicolumn{4}{c}{(c) Background replacement}
\end{tabular}
\\
&
\setlength{\tabcolsep}{.3mm}
\begin{tabular}{cccc}
\includegraphics[width=19mm]{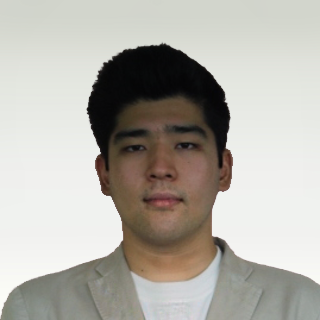}&
\includegraphics[width=19mm]{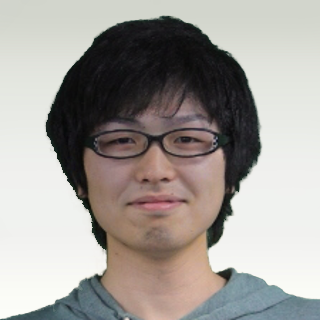}&
\includegraphics[width=19mm]{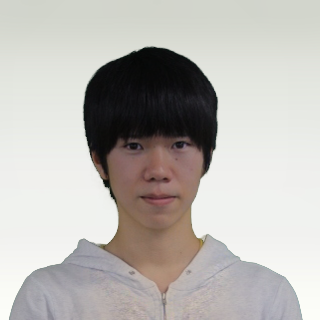}&
\includegraphics[width=19mm]{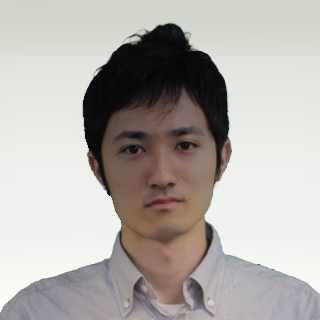}
\\
\multicolumn{4}{c}{(d) \cite{Shin2014}}
\end{tabular}
&
\setlength{\tabcolsep}{.3mm}
\begin{tabular}{cccc}
\includegraphics[width=19mm]{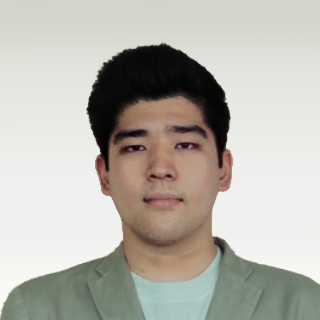}&
\includegraphics[width=19mm]{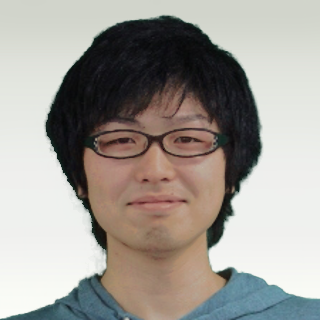}&
\includegraphics[width=19mm]{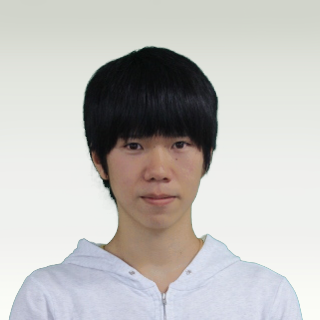}&
\includegraphics[width=19mm]{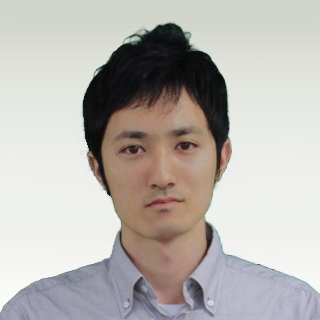}
\\
\multicolumn{4}{c}{(e) Ours}
\end{tabular}
\end{tabular}
\caption{Comparison with the background replacement result and \cite{Shin2014}. (a) Target facial skin color image, (b) Original images, (c) The background replacement result as a example, (d) The style transfer result by \cite{Shin2014}, and (e) The yearbook style photo using our algorithm.}
\label{fig:comparison_yearbook}
\end{figure*}

Figure~\ref{fig:result_compare1} compares our method with supplement methods \cite{Levin2004,JBU}. For the methods, we process each RGB color layer because they are proposed for color components in YUV color space.
Since colorization \cite{Levin2004} just spreads colors to boundaries, the result does not have the details of the original image.
Joint Bilateral Upsampling \cite{JBU} computes pixel values in hole regions by joint-filtering. We can see a contrast reduction in the top and the middle rows in Fig.~\ref{fig:result_compare1}.
The blue arrows and the red arrows indicate artifacts of each method.
Meanwhile our guide image filtering outputs sharp details of the original image while keeping the guide image color.

\begin{figure*}[t]
\centering
{\small
\setlength{\tabcolsep}{1mm}
\begin{tabular}{ccccc}
\includegraphics[width=35mm,clip]{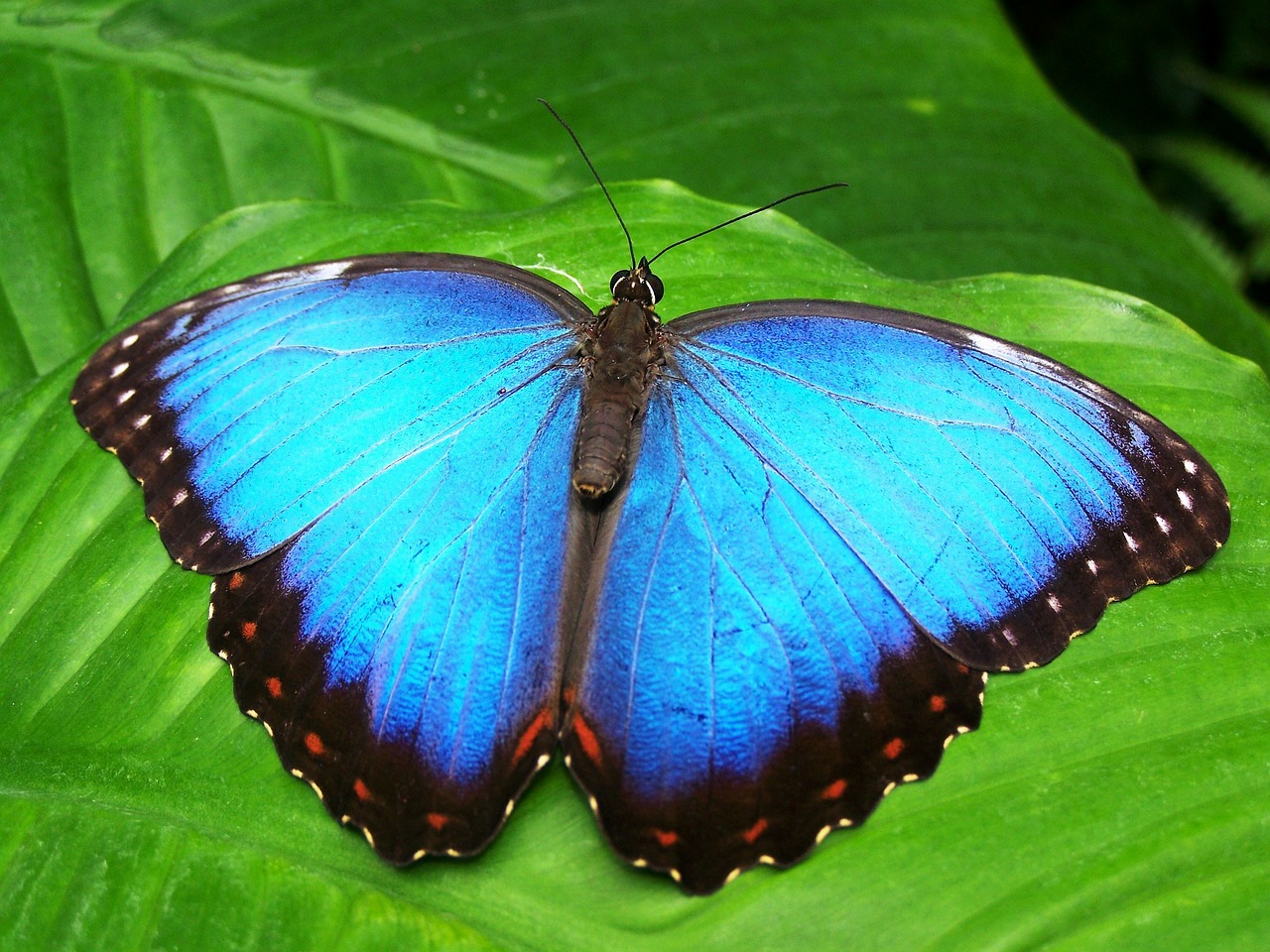} &
\includegraphics[width=35mm,clip]{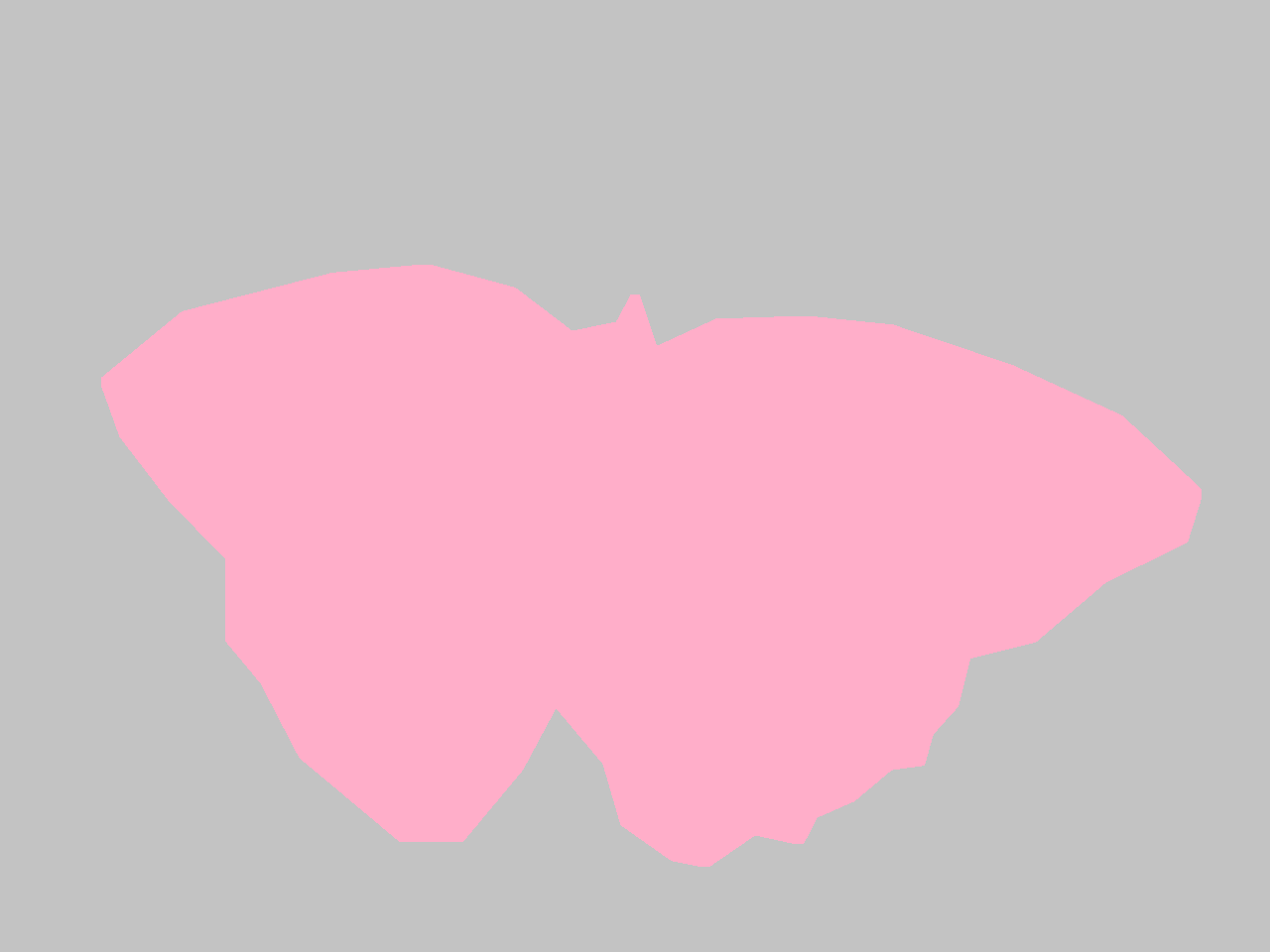} &
\includegraphics[width=35mm,clip]{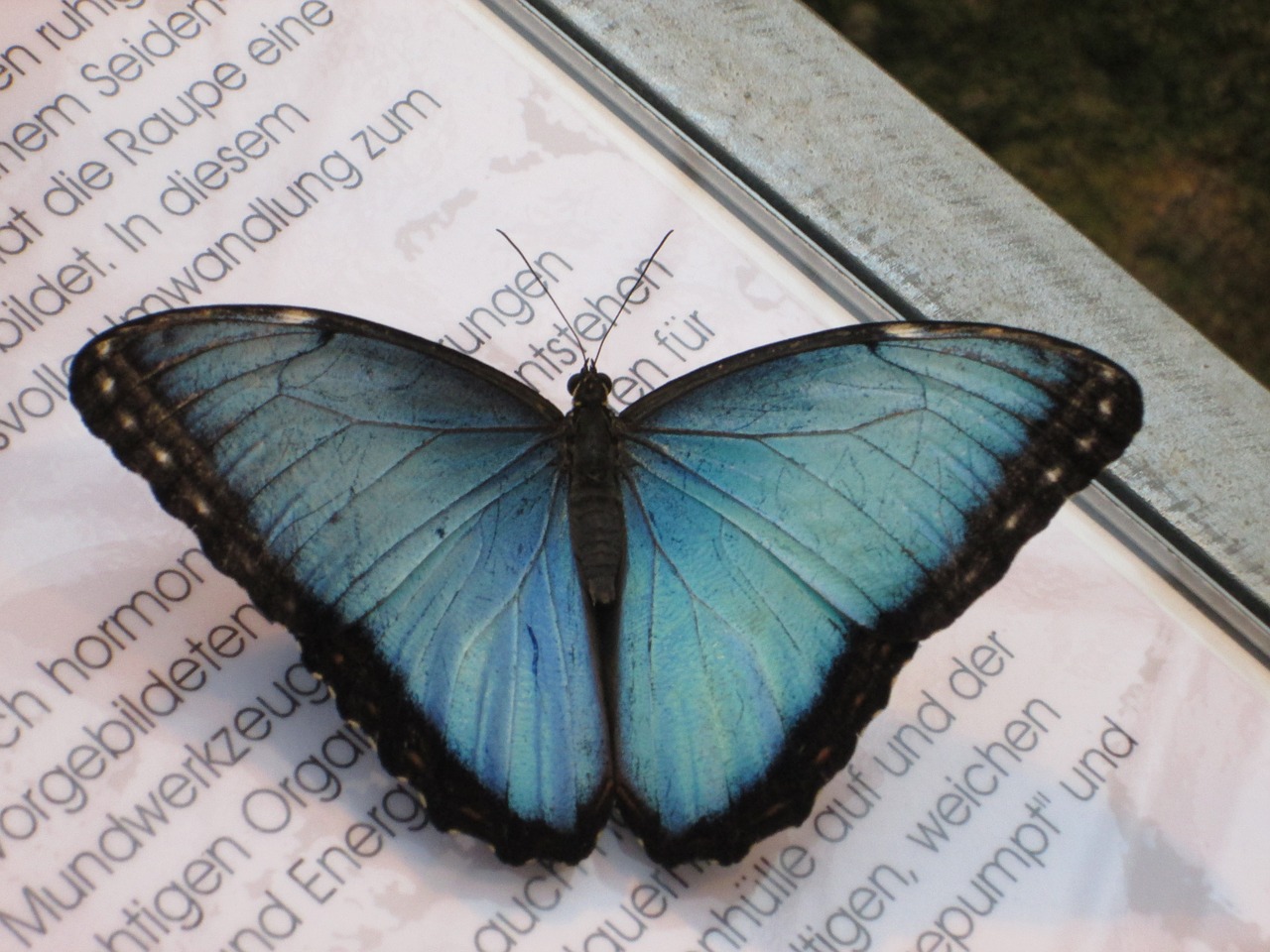} &
\includegraphics[width=35mm,clip]{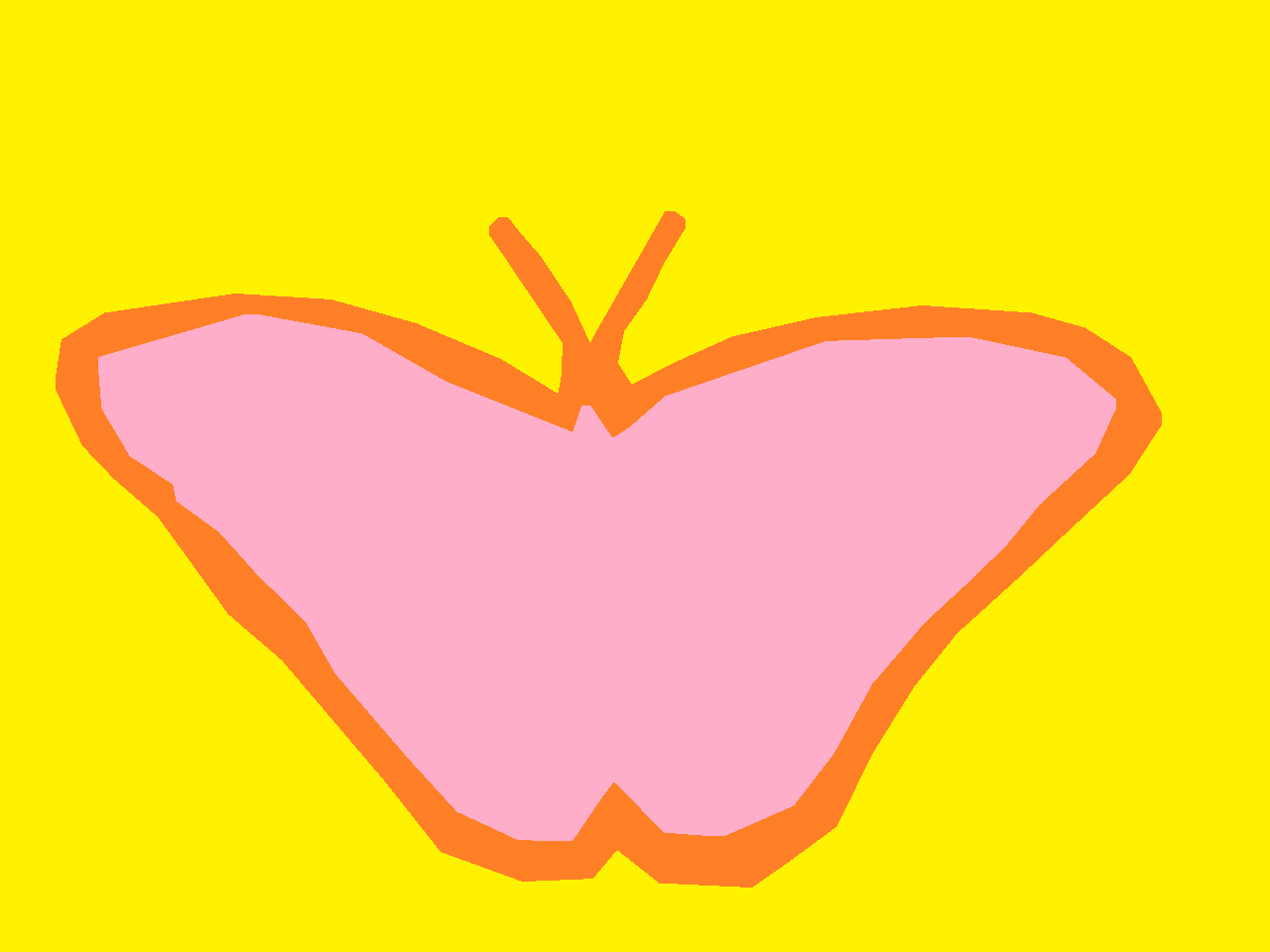} &
\includegraphics[width=35mm,clip]{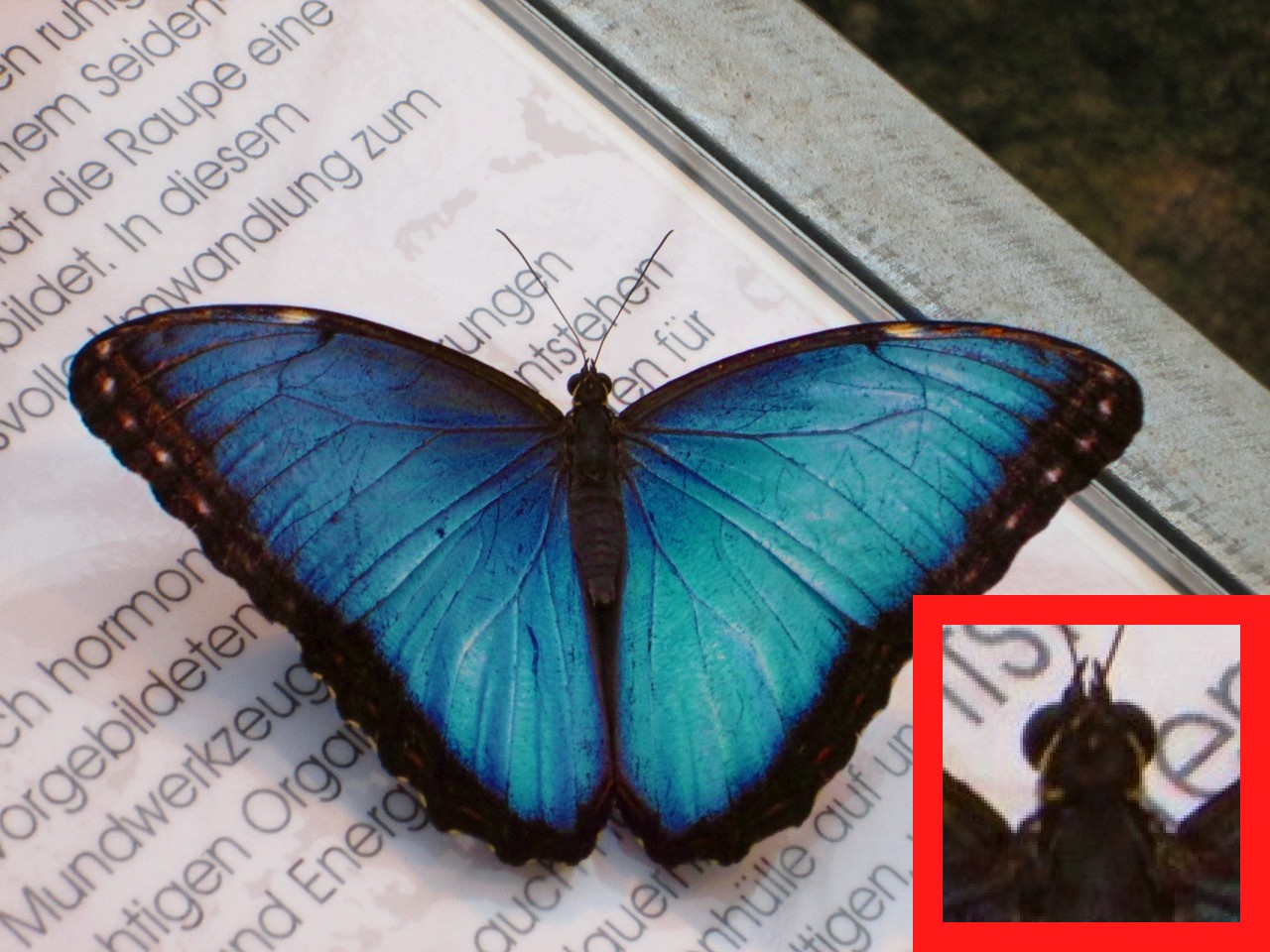} 
\\
\includegraphics[width=35mm,clip]{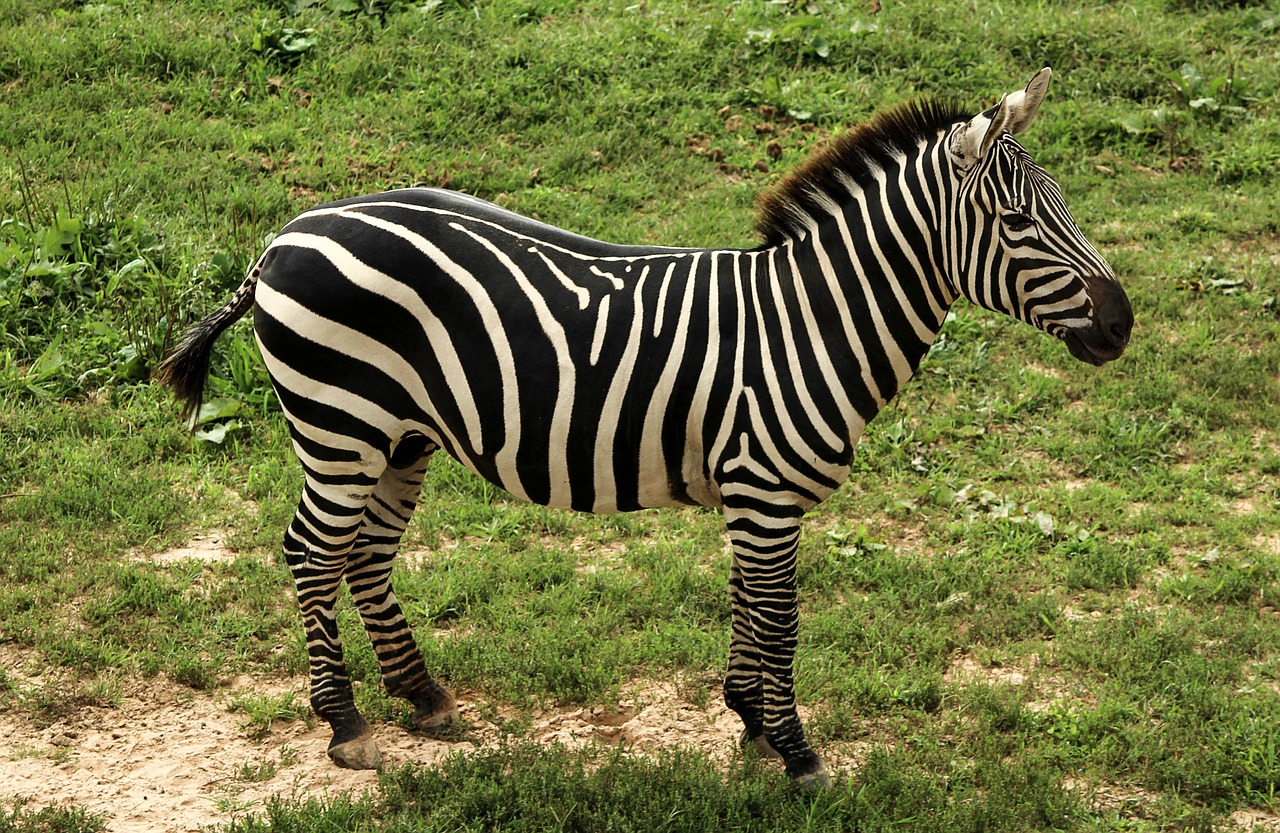} &
\includegraphics[width=35mm,clip]{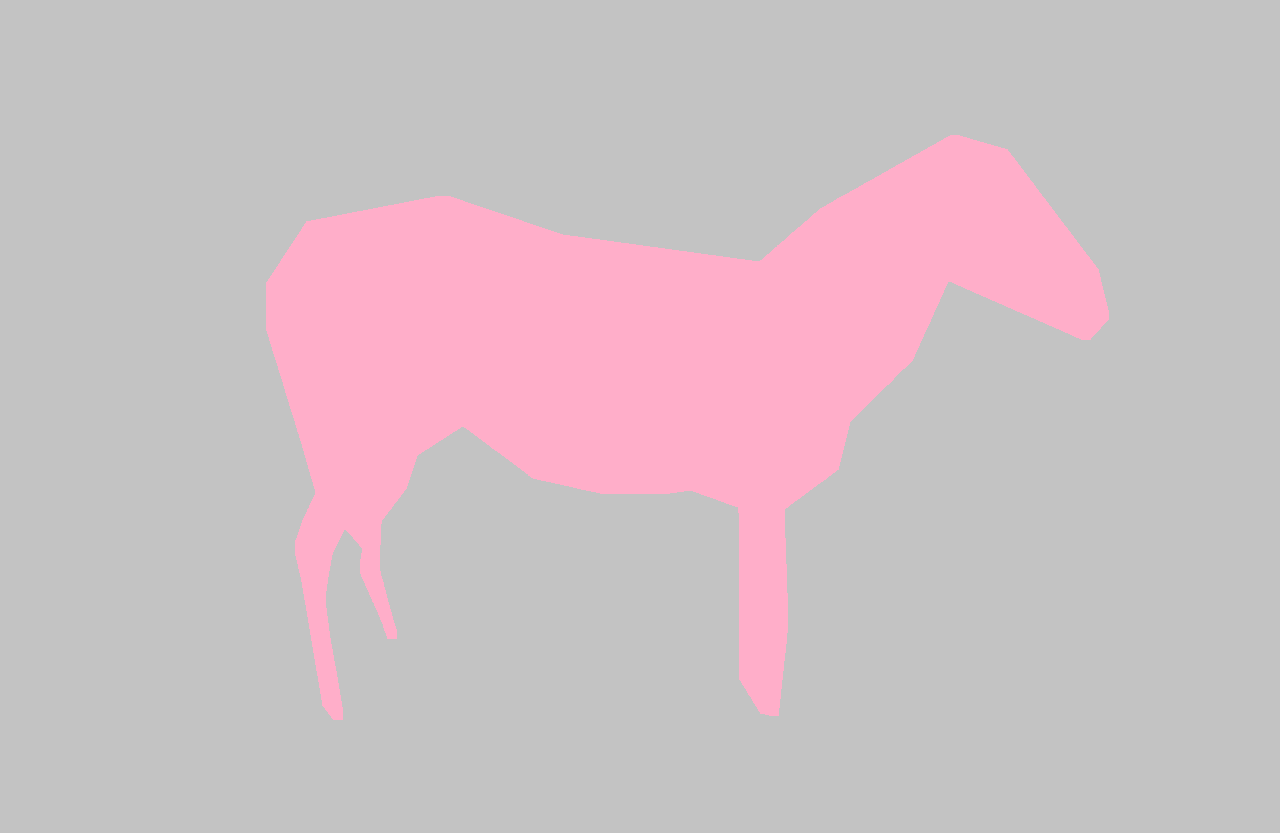} &
\includegraphics[width=35mm,clip]{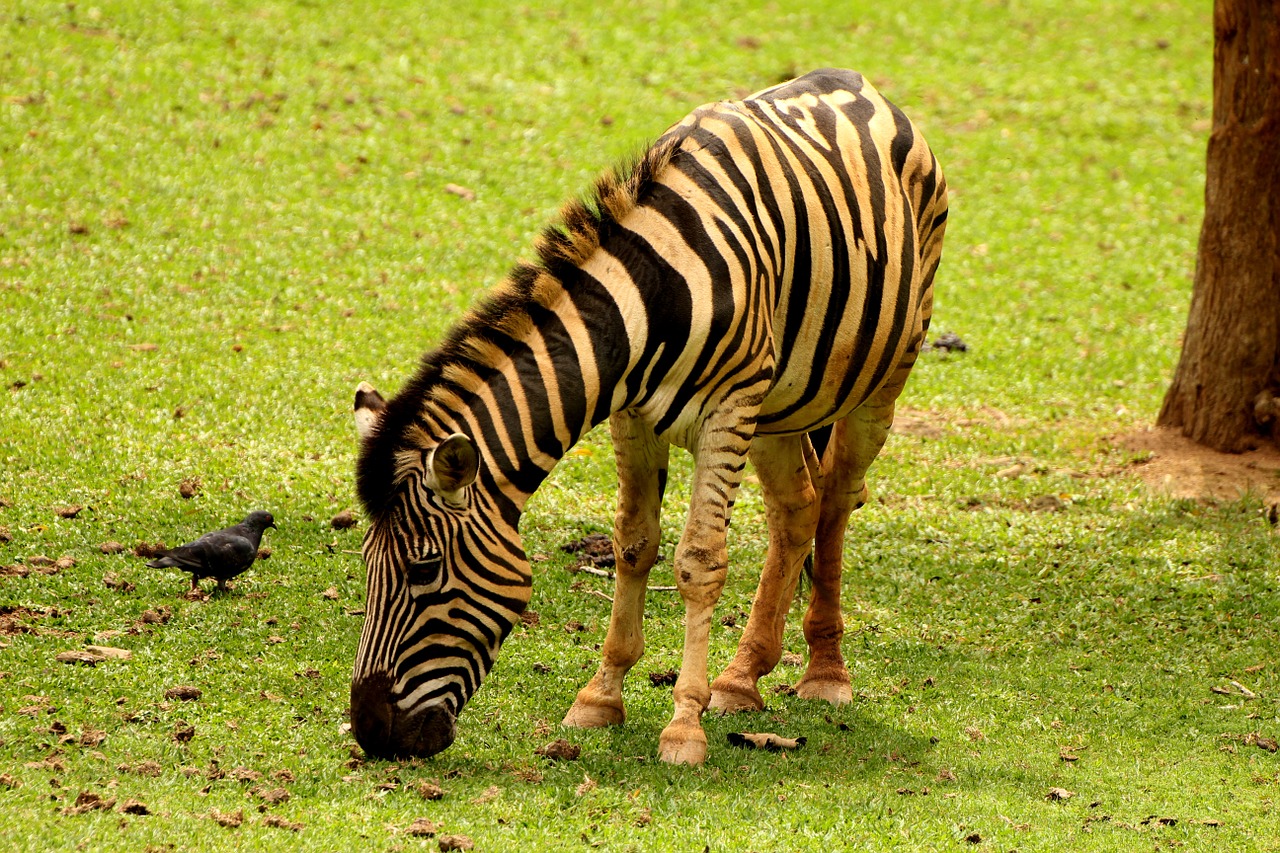} &
\includegraphics[width=35mm,clip]{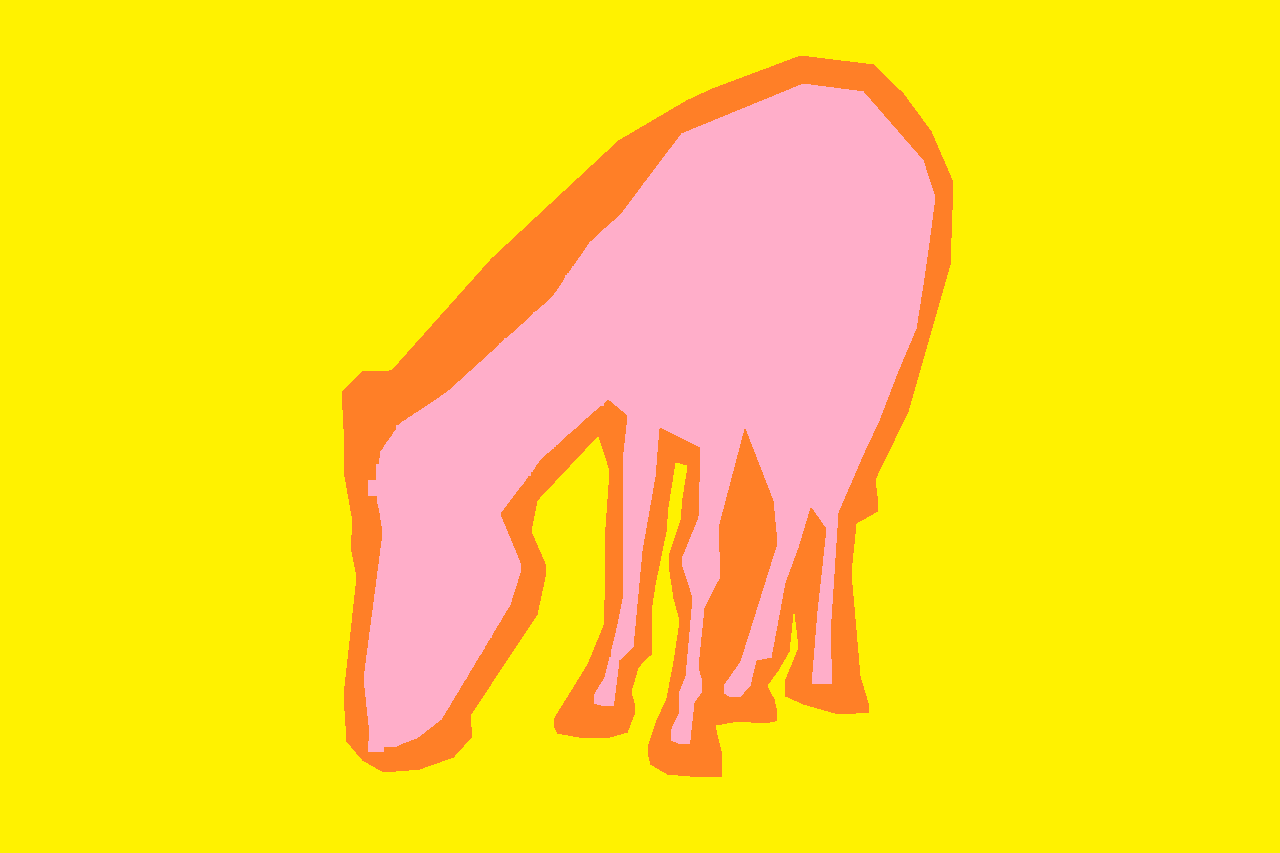} &
\includegraphics[width=35mm,clip]{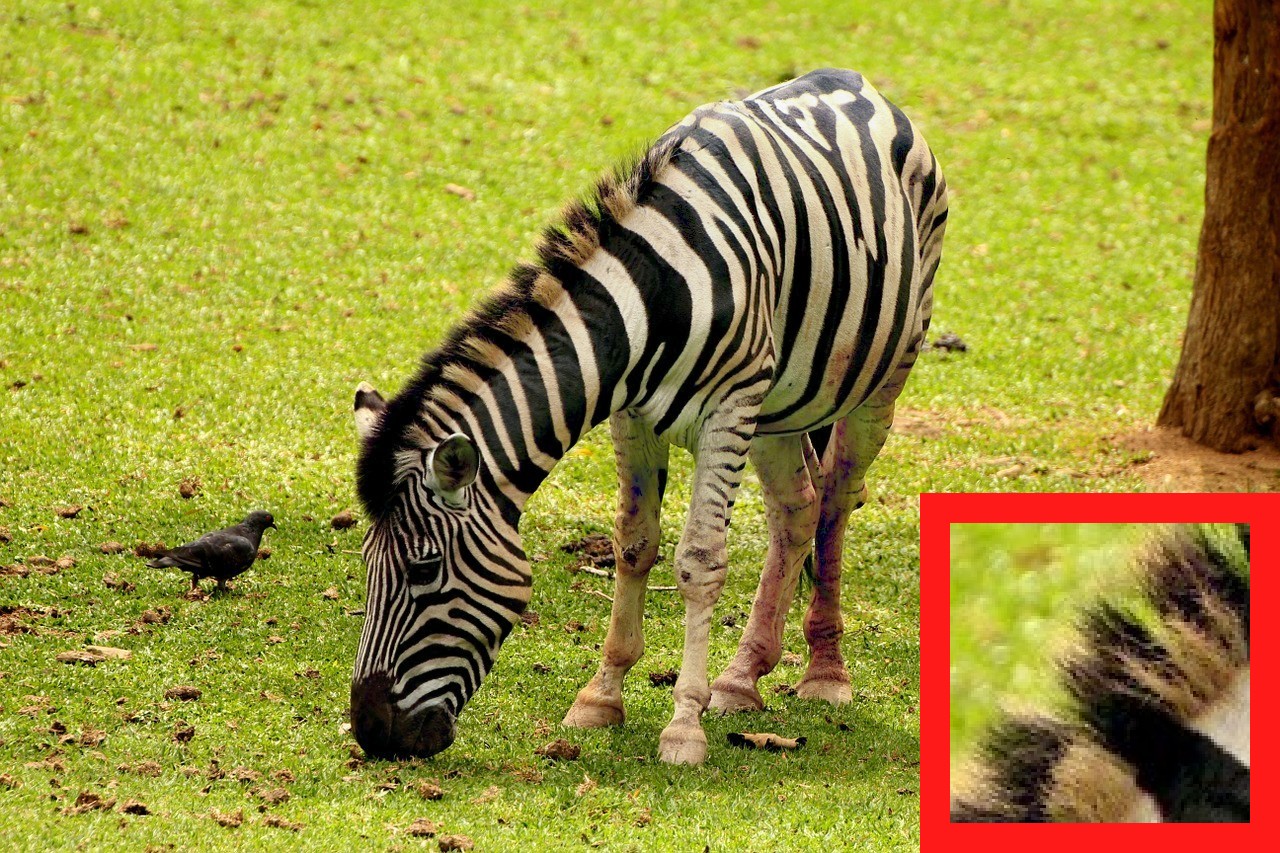} 
\\
\includegraphics[width=35mm,clip]{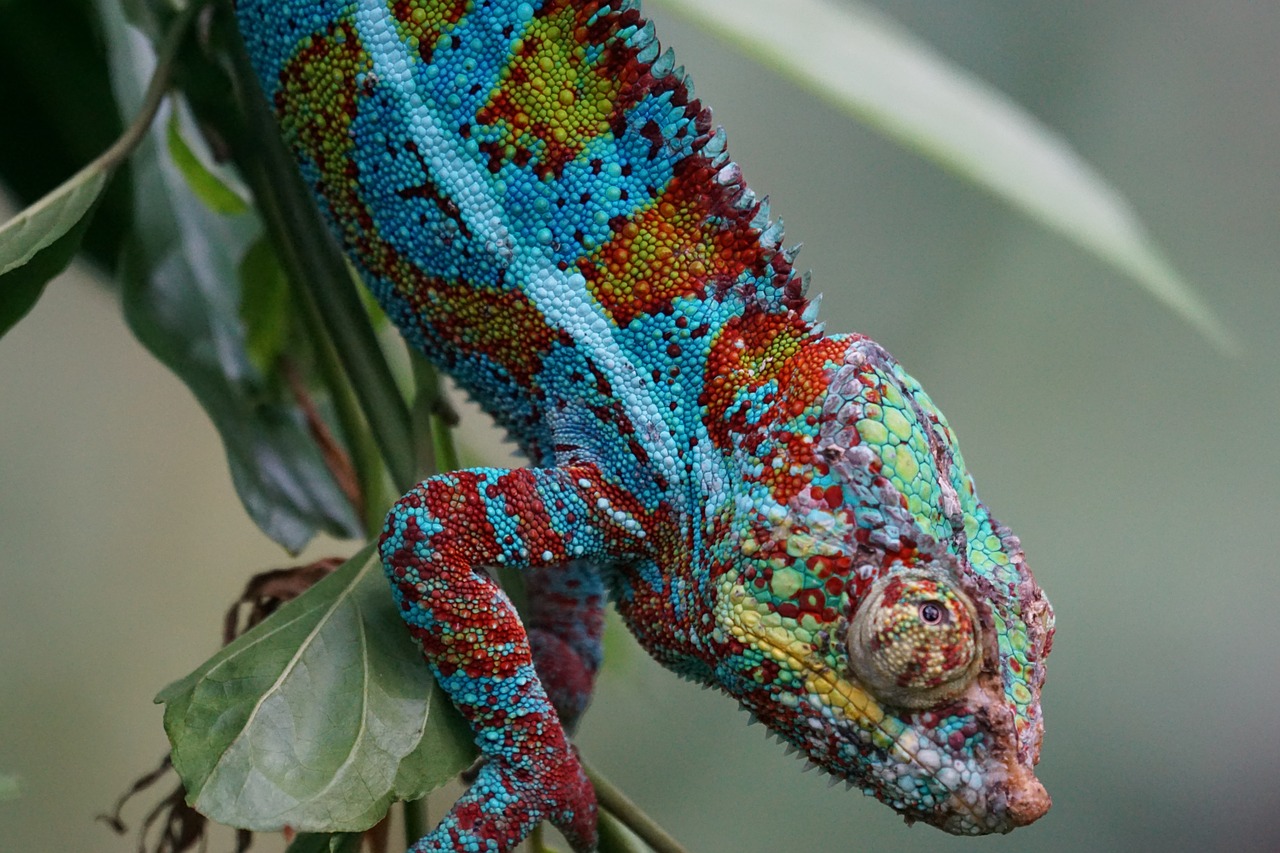} &
\includegraphics[width=35mm,clip]{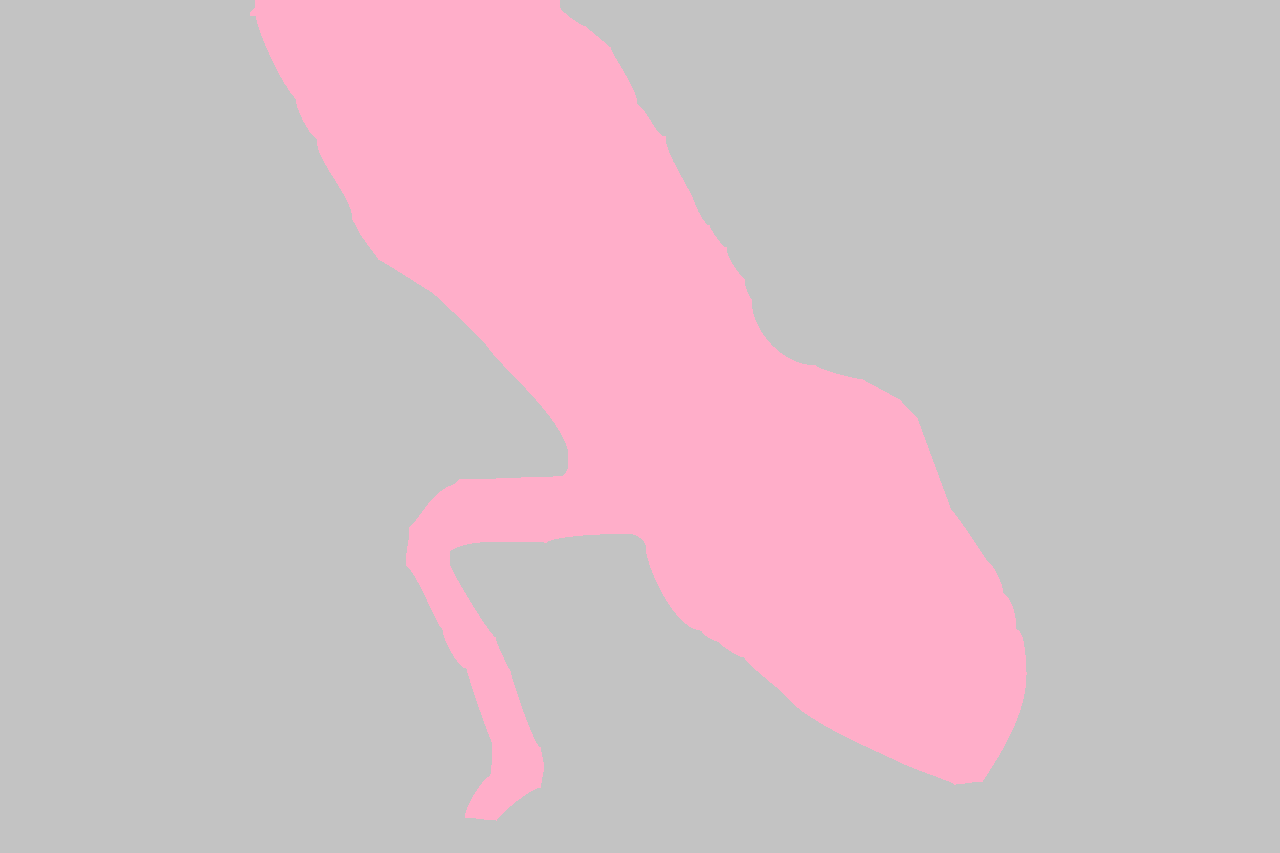} &
\includegraphics[width=35mm,clip]{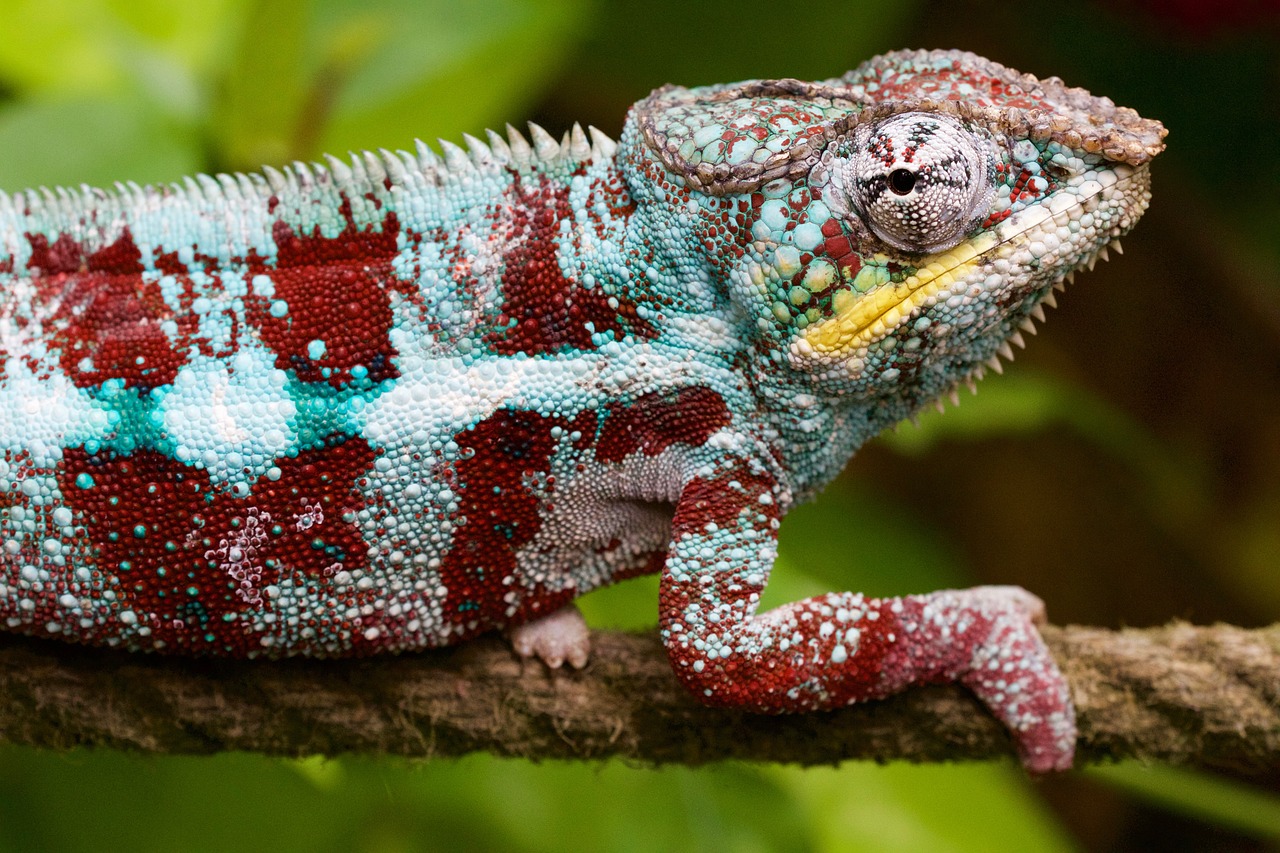} &
\includegraphics[width=35mm,clip]{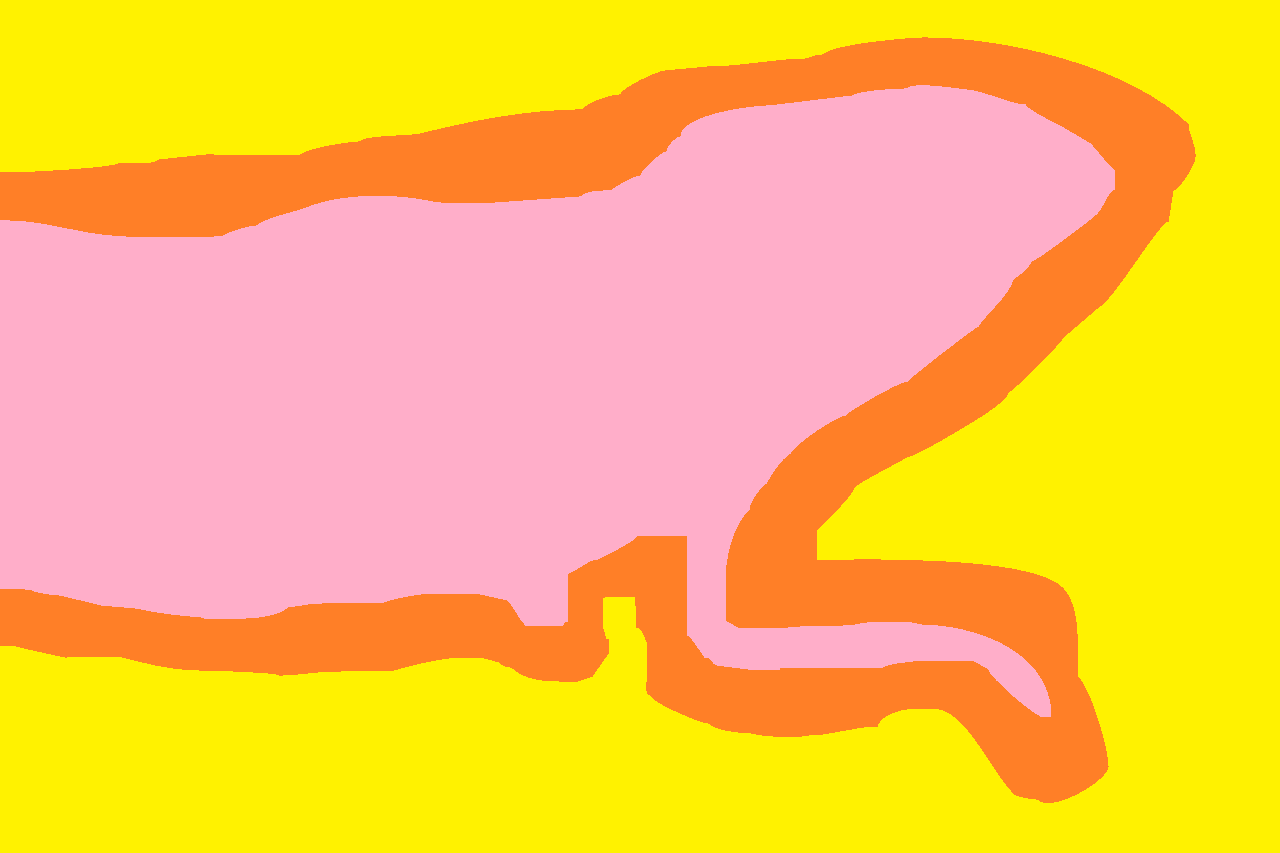} &
\includegraphics[width=35mm,clip]{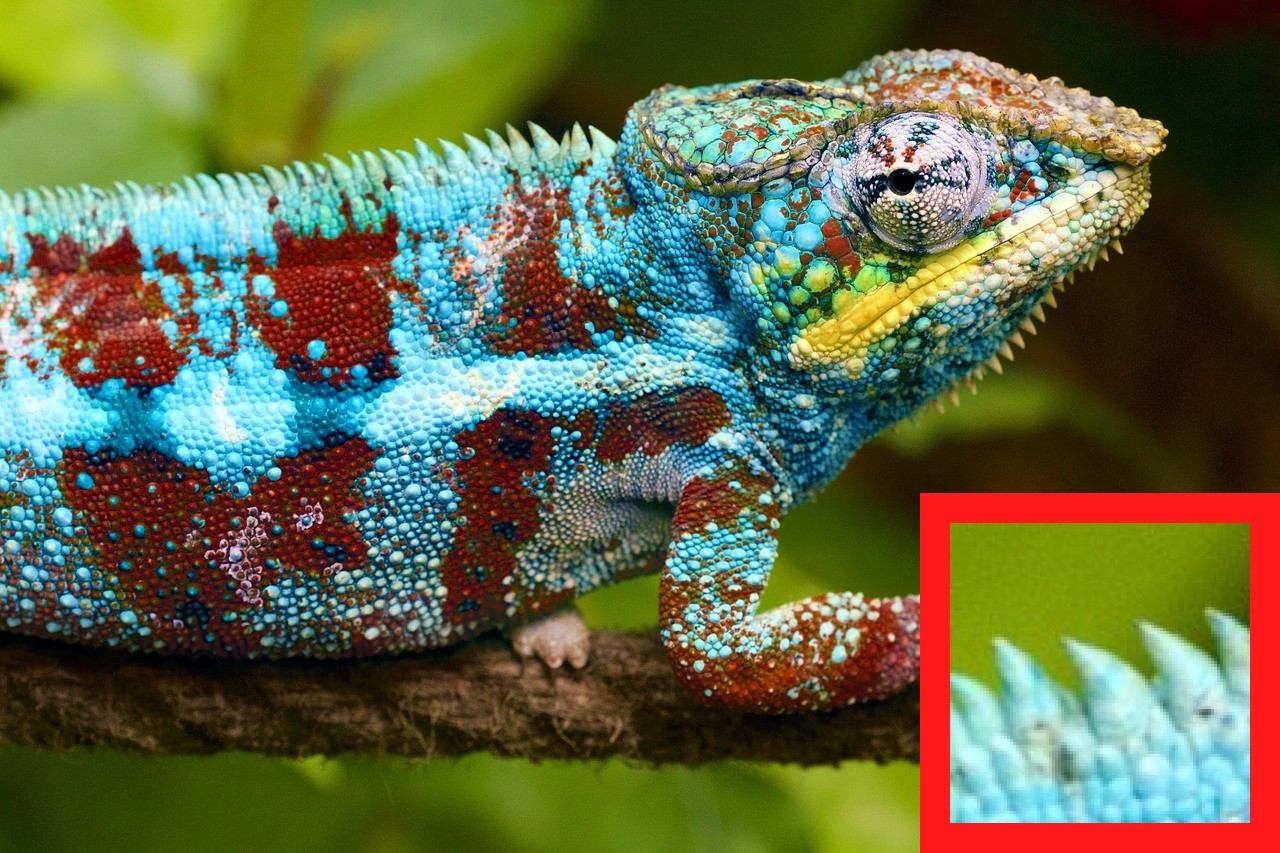} 
\\
(a) Target & (b) Each region & (c) Source  & (d) Each region & (e) Result
\end{tabular}
}
\caption{Semi-automatic color correction. (a) Target, (b) Each region in target image, (c) Source, (d) Each region in source image, and (e) Result. Colors in (b) and (d) indicate each region as Fig.~\ref{fig:eachregi}. The source and target images are available at https://pixabay.com.}
\label{fig:more}
\end{figure*}

Figure~\ref{fig:result_compare} shows a comparison with existing color transfer methods. 
For the results (c), Piti{\'e} {\it et al}.'s method \cite{Pitie2007} transforms the color distribution of the input image to the target image one, then gain noise artifacts of the color transformed input image are removed by Rabin {\it et al.}'s method \cite{Rabin2010}.
The gain noise artifacts can be seen in \cite{Pitie2007}.
In the results of the color grading \cite{Rabin2010} with\cite{Pitie2007}, the face and the background of the resultant image hava a similar color since this method is a uniform color transfer method. In addition, this method takes a long time because it needs iterative bilateral filtering.
In the non-rigid dense correspondence (NRDC) method \cite{HaCohen2011} and Jaesik {\it et al.} \cite{Jaesik2016}, the improvement in facial color is small, and the regions around the clothes are discolored. 
However, in our method (e), color correction is successfully done over the entire face, and look more natural.
From the above results, the proposed process gives satisfactory results despite the automatic processing.

Figure~\ref{fig:comparison_yearbook} shows a comparison with a background replacement result of the original image and \cite{Shin2014}.
The background replacement result has a distorted facial skin color due to the background color. 
Shin \textit{et al.}'s method transfers a style from an input image to a target image. Although the facial skin color of the result is almost the same as the target image one, this method cannot correct skin color of a person wearing glasses to the skin color of the target image. Our algorithm generates an image which has the target facial skin color even if a person wears glasses.

%
%

\subsection{Semi-automatic color correction}\label{sec:limit}

Our method can correct other images, such as animal photographs.
Our color correction method requires some regions, which are $\Omega_{S}$ as a foreground and $\Omega_{B}$ as a background.
Many methods have been proposed to detect face regions, but accurate object detection in natural scenes is still a challenging problem.

For color correction in natural scenes such as Fig.~\ref{fig:more}(c), we make each region manually as (b) and (d), where the colors indicate the region corresponding to Fig.~\ref{fig:eachregi}, respectively. Finally, our method adjusts the object color of the source image to the target one automatically.
Figure\ref{fig:more}(e) shows natural scene color correction results and the red box shows that our method corrects the colors of main object without boundary artifacts.
By implementating an automatic object detection method, this application will be an auto-application.

\begin{figure}[t]
\centering
{\footnotesize
\setlength{\tabcolsep}{1mm}
\begin{tabular}{ccc}
\includegraphics[width=25mm]{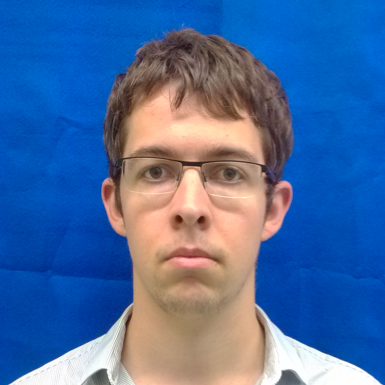} &
\includegraphics[width=25mm]{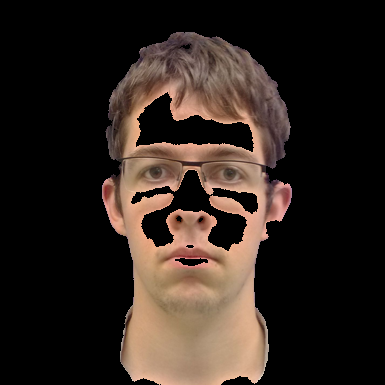} &
\includegraphics[width=25mm]{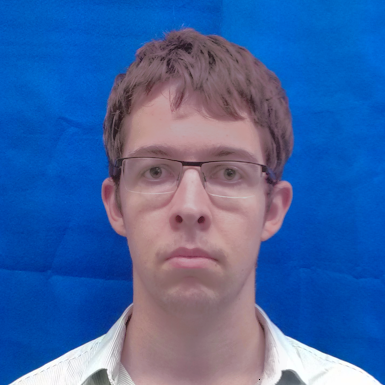} \\
\includegraphics[width=25mm]{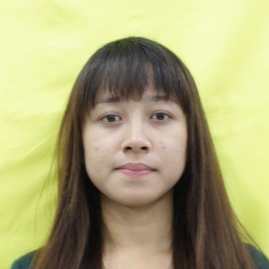} &
\includegraphics[width=25mm]{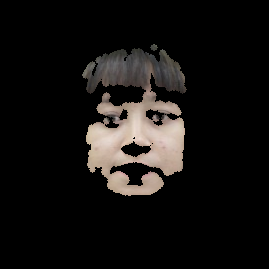} &
\includegraphics[width=25mm]{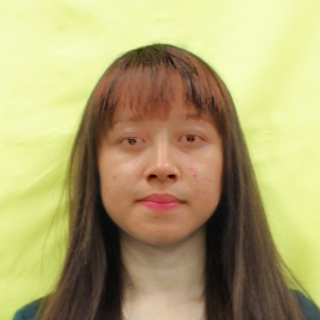} \\
(a) Original images & (b) Extracted skin & (c) Result
\end{tabular}
}
\caption{Bad results of our method due to the skin color extraction phase. In the original image the face and hair are a similar color.}
\label{fig:bad2}
\end{figure}
%

%
%

\subsection{Limitation}\label{sec:limit}

Figure~\ref{fig:bad2} shows example images where our method does not work well.
The face and hair are a similar color, hence, the facial skin color extraction process gets regions other than the face region (Fig.~\ref{fig:bad2}(b)).
As a result, our method outputs an image with a similar color between the face and hair (Fig.~\ref{fig:bad2}(c)).
Actually, the facial skin color extraction process seldom gets other regions even if the subject does not have white skin, so the guide image filtering results in the same color between the face and the other extracted regions.

%
%

\section{Conclusions}
\label{sec:conclusion}
This paper presents a guided facial skin color correction method.
The color grading method used in the procedure can be thought of as a correction method using a target (guide) image in a color space. On the other hand, our GIF can be thought of as a correction method using a guide image in an image space.
In future work, we will consider combining both methods to extend the range of constraint expressions to color correction so as to obtain more natural results.

\appendix

\section*{Appendix A: Fore/Background segmentation by matting} \label{sec:matting}

\begin{figure}[t]
\centering
{\small
\setlength{\tabcolsep}{1mm}
\begin{tabular}{ccccc}
\includegraphics[height=16mm]{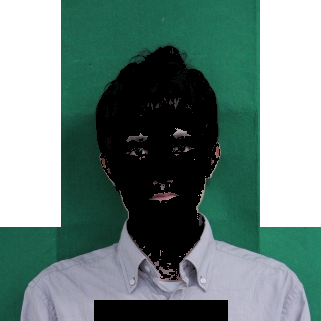} &
\includegraphics[height=16mm]{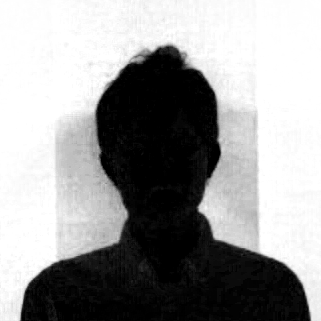} &
\includegraphics[height=16mm]{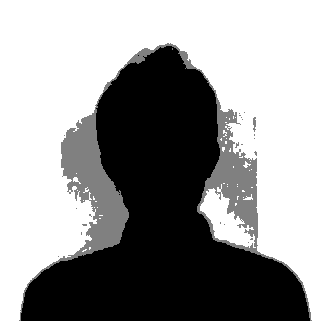} &
\includegraphics[height=16mm]{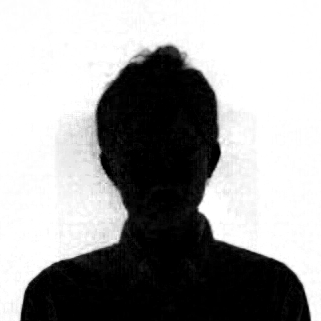} &
\includegraphics[height=16mm]{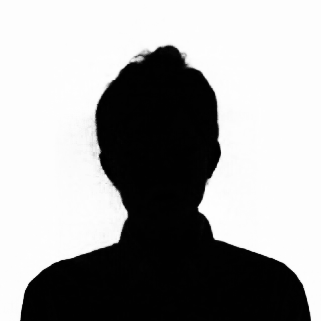}
\\
(a) & (b) & (c) & (d) & (e)
\end{tabular}
}
\caption{Image matting with region growing.}
\label{fig:matting}
\end{figure}

To obtain the alpha-mats that give natural blending results, we employ a \textit{closed form matting} method \cite{He2010} because the algorithm is also used in Sec.~\ref{sec:guide} and can be diverted. The method is based on continuous optimization in which labels are obtained as real numbers (soft labels) in the range of $[0,1]$.
However this method requires semantics (user assisted information) for fore/background regions around their boundaries. In order to avoid user assistance,
we also employ a region growing method used in an earlier matting method \cite{Sun2004} and perform matting iteratively.
The details are shown in Fig.~\ref{fig:matting} and described as follows:
\begin{enumerate}
\renewcommand{\labelenumi}{(\alph{enumi})}
\setlength{\parskip}{1mm}
\setlength{\itemsep}{0mm}
\item As the initial foreground $\Omega_F$, in addition to the skin color region $\Omega_{S}$, the hair region above the face (a black large region is roughly selected), and the clothes region below the face are used.
As the initial background $\Omega_B$, two rectangular regions on the left and right sides of the face are used.
\item Matting \cite{He2010} with the preconditioning conjugate gradient method is performed. A soft label $\alpha_i \in [0,1]$ is obtained at each pixel.
\item The pixels that are strongly regarded as foreground or background are added to the initial region, which is generated in (a), for the next iteration:\\ $\Omega^+_B := \{ i \ | \ \alpha_i \le 0.2  \}$, $\Omega^+_F := \{ i \ | \ 0.8 \le \alpha_i \}$ and then\\ $\{\alpha_i := 0 \ | \ i \in \Omega_B \cup \Omega^+_B\}$,  $\{\alpha_i := 1 \ | \ i \in \Omega_F \cup \Omega^+_F$\}.
\item Steps (b) and (c) are repeated a few times (4 times, in our experiment). The radius of the window size is halved to implement a coarse-to-fine approach. 
\item To reduce neutral colors and enhance them to be close to 0 or 1, a \textit{sigmoid function} is applied to the alpha-mat:\\ $\alpha_i := (1 + \exp( -10(\alpha_i-0.5)))^{-1}$.
\end{enumerate}

\bibliographystyle{plain}
\bibliography{myref}

\end{document}